\documentclass[10pt,superscriptaddress,aps,pra,twocolumn,showpacs,floatfix]{revtex4-2}

\usepackage[utf8]{inputenc}
\usepackage{graphicx}
\usepackage{amsmath,amsfonts,amssymb}
\usepackage{color}
\usepackage{epstopdf}
\usepackage{float}
\usepackage{physics}
\usepackage{placeins}
\usepackage{lipsum}
\usepackage{mathtools}
\usepackage{comment}
\usepackage{bbm}
\usepackage{soul}
\usepackage{xcolor}
\usepackage[normalem]{ulem}

\usepackage{silence}
\usepackage[colorlinks=true, allcolors={blue}]{hyperref}
\usepackage{nameref}

\newcommand{\orcid}[1]{%
  \href{https://orcid.org/#1}{\includegraphics[width=7pt]{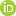}}%
}

\begin{document}

\preprint{APS/123-QED}

\title{Weak-to-Strong Measurement Transition with Thermal Instabilities: From Anomalous Amplification to Metrological Sensitivity}

\author{Marcos V. S. Lima~\orcid{0009-0000-2594-2713}}
\affiliation{Departamento de F\'{i}sica, Universidade Federal do Piau\'{i}, Campus Ministro Petr\^{o}nio Portela, CEP 64049-550, Teresina, PI, Brazil}

\author{Carlos H. S. Vieira~\orcid{0000-0001-7809-6215}}
\affiliation{Centro de Ci\^{e}ncias Naturais e Humanas, Universidade Federal do ABC,
Avenida dos Estados 5001, 09210-580 Santo Andr\'e, S\~{a}o Paulo, Brazil}

\affiliation{Department of Physics, State Key Laboratory of Quantum Functional Materials,
and Guangdong Basic Research Center of Excellence for Quantum Science,
Southern University of Science and Technology, Shenzhen 518055, China}

\author{Irismar G. da Paz~\orcid{0000-0002-9613-9642}}
\affiliation{Departamento de F\'{i}sica, Universidade Federal do Piau\'{i}, Campus Ministro Petr\^{o}nio Portela, CEP 64049-550, Teresina, PI, Brazil}

\author{Pedro R. Dieguez~\orcid{0000-0002-8286-2645}}
\affiliation{Department of Physics, Federal University of Paran\'{a}, P.O. Box 19044,  81531-980 Curitiba, Paran\'{a}, Brazil}

\author{Lucas S. Marinho~\orcid{0000-0002-2923-587X}}
\email{lucas.marinho@ufpi.edu.br}
\affiliation{Departamento de F\'{i}sica, Universidade Federal do Piau\'{i}, Campus Ministro Petr\^{o}nio Portela, CEP 64049-550, Teresina, PI, Brazil}


\begin{abstract}
Quantum measurement is physically realized through a finite dynamical interaction between a system and a measuring apparatus, giving rise to a continuous transition from weak to strong regimes. While this crossover is well understood under ideal conditions, the combined role of thermal instabilities and pre- and post-selection open dynamics has not been systematically addressed.
Here, we develop a generalized open-system framework to analyze the weak-to-strong measurement transition in the simultaneous presence of environmental decoherence and thermal noise. We model the probe as a thermal Gaussian state, explicitly incorporating temperature-dependent fluctuations in the measuring device, and include open-system evolution of the measured system prior to post-selection. By deriving the apparatus's final state, we show that the measurement statistics are modified in a nontrivial, highly sensitive manner by the temperature regime of the system’s thermal instabilities, the probe's thermal properties, and the particular choice of pre- and post-selection. This approach allows us to characterize how thermal effects reshape the weak-value condition, the anomalous amplification, and the resulting metrological sensitivity, demonstrating the protocol's practical utility for precision measurements across the full measurement crossover.
\end{abstract}

\maketitle

\section{INTRODUCTION}

The quantum measurement process should not be understood solely in terms of its two limiting regimes, weak~\cite{AharonovAlbertVaidman1988PRL} and projective~\cite{vonNeumann1955book,Ozawa1984JMP}, but rather as a continuous transition connecting them. In this broader perspective, the weak-to-strong~\cite{Pan2020Nature,TurekPRA2015,Orszag2021PRA,Turek2023PLA,Turek2023EPJP,Turek2024PRA,Turek2026Arxiv,dieguez2018information,PhysRevA.111.012220,lustosa2025emergence} measurement transition provides a fundamental framework for understanding how information is extracted from a quantum system as the measurement strength interpolates, and how disturbance, back-action, and pre- and post-selection~\cite{Wu2011PRA} jointly determine observable outcomes. Beyond its conceptual relevance to quantum information, this transition also bears direct thermodynamic implications, as measurement back-action can modify a system’s internal energy and enable the design of measurement-powered thermal devices~\cite{yi2017single,brandner2015coherence,lisboa2022experimental,dieguez2023thermal,VIEIRA2023100105,PhysRevA.109.042424,lisboa2026correlations,bv4w-jr6q}. In this light, understanding how measurement regimes interpolate from weak to strong provides not only a unified perspective on quantum information acquisition but also valuable insights into their energetic implications~\cite{elouard2017role,latune2025thermodynamically,6p2c-nbg5}.

In particular, the weak measurement regime corresponds to the limit of vanishing interaction strength, where the disturbance of the system is minimal, and the measurement outcome yields only a small correction to the detection probability~\cite{DresselRevMPhys2014}. In this limit, the weak value naturally emerges, being identified as the relative correction to a detection probability due to a small (weak interaction regime) intermediate perturbation~\cite{DresselRevMPhys2014}. An important property of weak values is their potential to assume values outside the eigenvalue spectrum of the measured observable. In particular, when the initial and final post-selected states are nearly orthogonal, the weak value can become anomalously large, as first emphasized in the seminal Aharonov-Albert-Vaidman work~\cite{AharonovAlbertVaidman1988PRL}. Recently, the weak value technique has been used to generate non-Gaussian states with high fidelity~\cite{Yao2026npjQuantumInformation}, to map the ultrafast temporal dynamics of ultra-short laser pulses into measurable macroscopic shifts~\cite{Sahoo2025Nanoscale}, to achieve high sensitivity based on preselection-free weak measurements~\cite{Zifu2026COP}, to extend quantum metrology beyond the weak interaction limit~\cite{Yoo2025QST}, and to precisely detect attosecond time delays~\cite{HuangPRA2026}.

While weak values arise in the weak disturbance limit, increasing the measurement strength progressively modifies the system–meter correlations and the associated detection statistics. As the interaction becomes stronger, the measurement outcome gradually approaches the standard expectation associated with projective measurements, and the possibility of anomalous amplification diminishes~\cite{DresselRevMPhys2014}. Weak values should therefore be regarded as limiting cases within a more general framework of quantum measurement, rather than as isolated quantities. The weak-to-strong transition naturally provides the context for examining how weak-value behavior continuously transforms into conventional eigenvalue statistics, and how post-selection influences measurement outcomes across varying disturbance regimes~\cite{Pan2020Nature}.

When post-selection follows a pre-measurement interaction, the system is effectively open in the interval preceding the post-selection step, since conditioning on a final state makes the resulting statistics sensitive to uncontrolled environmental degrees of freedom. In this scenario, thermal instabilities acting prior to post-selection can significantly modify the final outcome of the quantum measurement. Indeed, in the context of the quantum switch, for instance, it has been shown that thermal fluctuations occurring before post-selection can qualitatively alter the generation of superpositions of causal orders~\cite{Dieguez2024CommPhys}. Remarkably, the resulting behavior was shown to depend sensitively on both the temperature regime of the instabilities and the particular post-selected state, revealing a nontrivial interplay between environmental noise and conditional quantum statistics. Furthermore, recent work has analyzed open dynamics occurring prior to the measurement interaction~\cite{Eliahu2025arxiv}, revealing noise-resilience advantages in the weak limit, but without fully accounting for temperature-dependent thermal effects. While the continuous crossover from weak to strong measurements has been systematically characterized in both experimental and theoretical frameworks~\cite{Pan2020Nature,TurekPRA2015,Orszag2021PRA,Turek2023PLA,Turek2023EPJP,Turek2024PRA,Turek2026Arxiv}, these foundational studies predominantly restrict their analyses to idealized closed systems or focus strictly on decoherence induced by the measuring apparatus. Consequently, the ubiquitous effects of thermal instabilities acting directly upon the target system throughout the measurement transition have remained largely unaddressed. Motivated by this open problem, we adopt a broader perspective here to investigate how thermal instabilities, taking place before post-selection, influence the complete weak-to-strong measurement transition, characterizing their impact on anomalous amplification and demonstrating the protocol's metrological usefulness for precision measurements.

In this work, we develop a dynamical description of the measurement process that explicitly accounts for both environmental decoherence of the pointer and thermal noise preceding post-selection. The model considers a continuous interaction in which the measuring device—the pointer—becomes progressively correlated with the microscopic system. The apparatus is described by a general thermal Gaussian state, providing a tunable probe whose temperature controls the strength of thermal fluctuations. Concurrently, the system evolves under open dynamics represented by a Generalized Amplitude Damping (GAD) channel applied prior to post-selection. This framework enables us to derive the apparatus's exact final state and determine how decoherence and thermal effects modify the measurement statistics across different interaction strengths.
Within this framework, measurement strength, environmental noise, and post-selection jointly determine the observed outcomes. Their interplay becomes particularly nontrivial throughout the weak-to-strong crossover: thermal fluctuations modify not only the pointer distribution, but also the conditions under which the weak-value regime is recovered. We demonstrate that the resulting behavior depends sensitively on both the temperature scale and the chosen post-selection, leading to qualitatively distinct regimes. In particular, we determine how the weak-value condition is modified by pre–post-selection noise, identify the regimes in which anomalous amplification persists or is suppressed, and clarify how the operational meaning of the measured quantity evolves as the interaction strength approaches the projective limit.

This work is structured as follows: In Sec.~\ref{sec:theory}, we expand the conventional weak measurement formalism beyond the limit of weak coupling and account for thermal instabilities. Therefore, in Sec.~\ref{sec:results}, we apply the Generalized Amplitude Damping (GAD) channel, and we explore how weak value anomalies and the success probability of the protocol are affected by thermal instabilities effects in Secs.~\ref{sec:weak_value} and \ref{sec:psucc_main}, respectively. The transition from weak to strong measurement is investigated in Sec.~\ref{sec:transition_quantification}, exhibiting generally non-monotonic behavior for non-classical pointer states. Finally, in Sec.~\ref{sec:Amplification}, the impact of the thermal bath on the measurement process is investigated through the anomalous shifts in the position and momentum of the pointer, and the metrological sensitivity to establish the protocol's usefulness for high-precision tasks. In Sec.~\ref{sec:discussion}, we discuss our results and possible implications.

\section{Theoretical Model}\label{sec:theory}

\begin{figure}[ht]
\centering
\includegraphics[scale = 0.11]{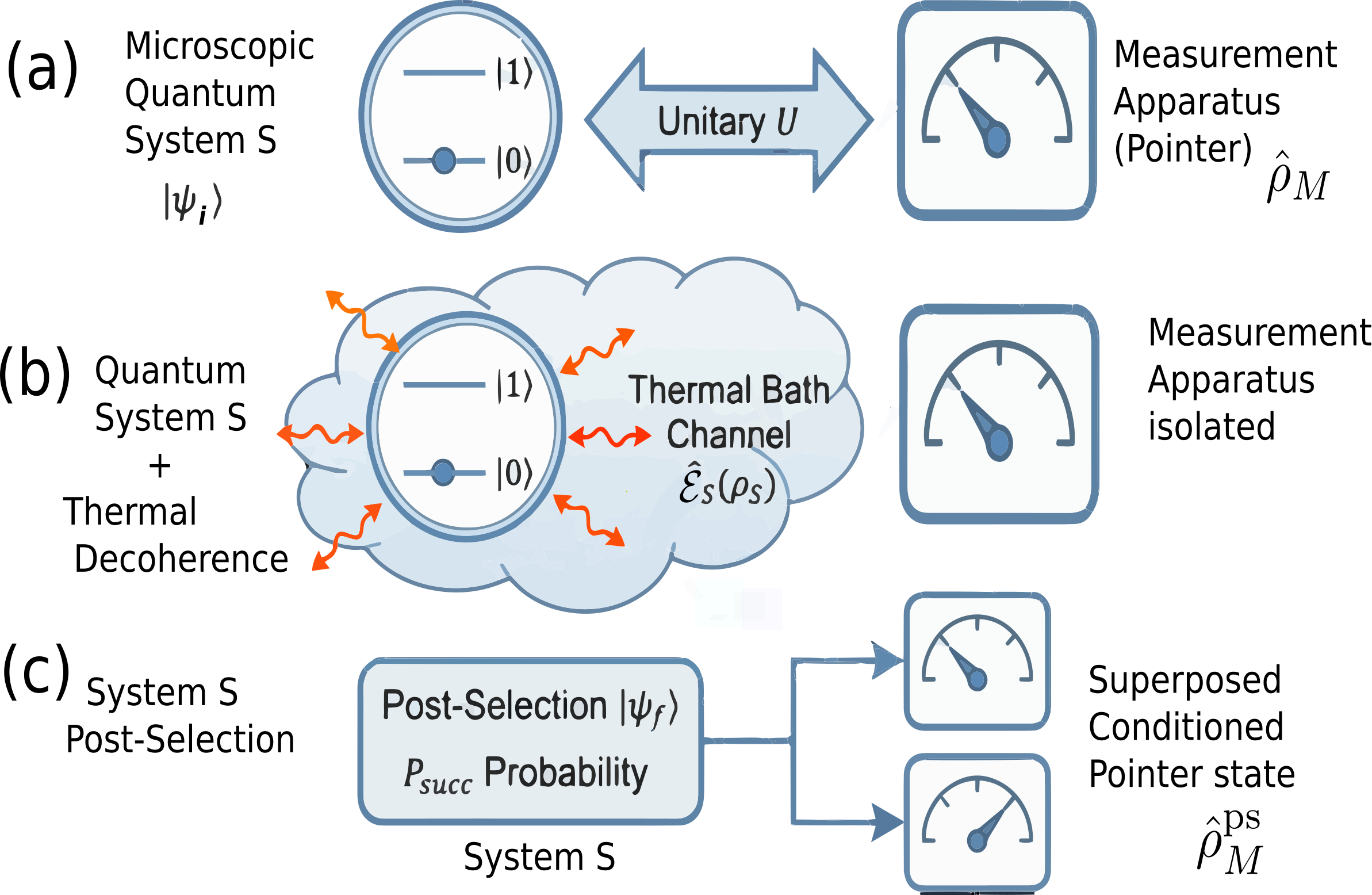}
\caption{Weak measurement scheme under thermal instabilities. (a) The microscopic two-level system $S$ interacts unitarily with the measurement apparatus (meter) $M$. (b) After the measurement interaction, only the system $S$ is subjected to a thermal quantum channel $\mathcal{E}_S$, while the meter remains isolated from the environment. (c) The system $S$ is post-selected in the state $|\psi_f\rangle$ with success probability $P_{\mathrm{succ}}$, resulting in a superposed conditioned pointer state $ \hat{\rho}_M^{\text{ps}}$ that carries the signature of the weak interaction and thermal signatures.}
\label{fig_model}
\end{figure}

In this section, we extend the standard weak-measurement formalism to account for environmental decoherence and thermal noise. We consider a scenario in which the microscopic system $S$ first undergoes a measurement interaction with the apparatus (meter) $M$, acting as a pointer [see Fig.~\ref{fig_model}(a)]. The standard von Neumann measurement coupling interaction is governed by the Hamiltonian $\hat{H} = g \delta(t-t_0) \hat{A} \otimes \hat{P}$, where the interaction is assumed impulsive at $t = t_0$~\cite{TurekPRA2015}. Here, $g$ is the coupling constant, and $\hat{P}=i/(2\sigma)(\hat{a}^{\dagger}-\hat{a})$ is the momentum operator conjugated to the position operator $\hat{X}=\sigma(\hat{a}^{\dagger}+\hat{a})$ (we assume $\hbar = 1$, henceforward), where $\sigma$ is the width of the initial Gaussian pointer state. For a system observable satisfying $\hat{A}^2 = \hat{\mathbbm{1}}_S$, where $\hat{\mathbbm{1}}_S$ represents the identity operator on the system's Hilbert space, the time evolution operator $\hat{U}$ can be appropriately decomposed into displacement operators $\hat{D}$ acting on the pointer space \cite{TurekPRA2015}
\begin{equation}\label{eq:Turek2015}
    \hat{U} = e^{-ig\hat{A}\otimes\hat{P}} = \frac{1}{2}( \hat{\mathbbm{1}}_S + \hat{A})\otimes \hat{D}\left(\frac{s}{2}\right) + \frac{1}{2}( \hat{\mathbbm{1}}_S - \hat{A}) \otimes \hat{D}\left(-\frac{s}{2}\right),
\end{equation}
where $s = g/\sigma$ encodes the measurement strength. In turn, the regimes $s \ll 1$ and $s \gg 1$ correspond to weak and strong system–pointer coupling, respectively.

The weak measurement protocol proceeds in two distinct stages. We assume the total system is initially prepared in the uncorrelated product state $\hat{\rho}(0) = \hat{\rho}_S \otimes \hat{\rho}_{M}$, where $ \hat{\rho}_S  = \ket{\psi_i}\bra{\psi_i}$ is the initial prepared state of the system being measured. The measurement apparatus is initialized in the most general bosonic single-mode Gaussian state. We employ a Gaussian pointer for its historical precedent, analytical tractability, and experimental feasibility. Conceptually, Gaussian wavepackets have served as the standard continuous-variable pointers since the seminal formulation of weak values~\cite{AharonovAlbertVaidman1988PRL}. Mathematically, they enable exact closed-form solutions even when the apparatus undergoes non-unitary thermal decoherence~\cite{serafini2017quantum}. Furthermore, they perfectly map onto the primary experimental platforms, seamlessly describing both the squeezed states in photonic systemns~\cite{LloydRevModPhys2012} and the motional ground states of trapped-ion architectures~\cite{Pan2020Nature}. Explicitly, this initial state is decomposed as
\begin{equation}\label{eq:rho_M_initial}
    \hat{\rho}_{M} = \hat{D}(\alpha) \hat{S}(\zeta) \hat{\rho}_{\mathrm{th}}(\bar{n}) \hat{S}^\dagger(\zeta) \hat{D}^\dagger(\alpha).
\end{equation}
Here, $\hat{D}(\alpha)$ is the displacement operator with complex amplitude $\alpha$, and $\hat{S}(\zeta)$ is the squeezing operator defined by the parameter $\zeta = r e^{i2\chi}$, where $r$ is the squeezing amplitude and $\chi$ is the squeezing phase~\cite{Olivares2012,LloydRevModPhys2012,MarianPRA1993}. Recently, the connection between this squeezing phase and position-momentum correlations has been explored within metrological protocols~\cite{OzielPRA2026,PortoEPJP2026,PortoPRA2025,Porto2024Scripta}, interferometry~\cite{MarinhoSciRep2024,MarinhoPRA2020,Oziel2019MPLA}, Gouy phase~\cite{ThiagoNJP2025}, quantum coherence~\cite{daSilva2024EPJP}, irrealism~\cite{LustosaPRA2020} and continuous-variable quantum computation~\cite{upreti2025interplayresourcesuniversalcontinuousvariable,upreti2025symplecticcoherencemeasurepositionmomentum}. The core of the decomposition is the thermal state $\hat{\rho}_{\mathrm{th}}(\bar{n})$, which is completely characterized by the mean thermal occupation number $\bar{n}= [\exp(\beta_M \omega_M)-1]^{-1}$. This occupation follows the Bose-Einstein statistics associated with the initial temperature $T_M$ of the apparatus, where $\beta_M=1/(k_B T_M)$ is the inverse thermal energy and $k_B$ is the Boltzmann constant. Immediately after the instantaneous von Neumann interaction, the state of the joint system is $\hat{\rho}_{\text{int}} = \hat{U} (\hat{\rho}_S \otimes \hat{\rho}_{M}) \hat{U}^\dagger$.
Following this rapid unitary step, and before post-selection, the system $S$ undergoes open quantum dynamics [see Fig.~\ref{fig_model}(b)] arising from its unavoidable coupling to the environment~\cite{JoosZeh2003book,schlosshauer2007decoherence,Marinho_2018EPL,Marinho_2023EPL}. In contrast, the measurement apparatus is modeled as a macroscopic mixed state, with decoherence effectively incorporated into its initial description. This characterization treats $M$ as a robust, quasi-classical entity possessing stable pointer states, in clear distinction to the fragility of the microscopic system $S$~\cite{Zurek_Review2003}. Such an approximation holds well provided the time interval preceding post-selection is significantly shorter than the pointer's coherence time. Under these conditions, the apparatus remains effectively isolated from environmental noise, evolving via the identity map, while $S$ undergoes decoherence effects. This intermediate evolution can be described by a general quantum channel $\mathcal{E}_S$ acting exclusively on the system, yielding the global density matrix $\hat{\rho}_{\text{dec}} = (\mathcal{E}_S  \otimes \hat{\mathbbm{1}}_M ) [\hat{\rho}_{\text{int}}]$.
Using the Kraus decomposition for the channel $\mathcal{E}(\cdot) = \sum_j \hat{K}_j (\cdot) \hat{K}_j^\dagger$~\cite{kraus1983states,nielsen2010quantum}, the final state before post-selection can be obtained
\begin{equation}\label{eq:rho_dec}
    \hat{\rho}_{\text{dec}} = \sum_j (\hat{K}_j \otimes \mathbbm{1}_M) \hat{U} (\hat{\rho}_S \otimes \hat{\rho}_{M}) \hat{U}^\dagger (\hat{K}_j^\dagger \otimes \mathbbm{1}_M).
\end{equation}

Proceeding with the standard steps of the weak-value protocol~\cite{AharonovAlbertVaidman1988PRL}, we post-select the system $S$ in the state $\ket{\psi_f}$ [see Fig.~\ref{fig_model}(c)]. The final reduced state of the apparatus $M$ is obtained by projecting the system onto $\ket{\psi_f}$ and tracing out the system degrees of freedom $\hat{\rho}^{\text{ps}} = \text{Tr}_S \left[ (\ket{\psi_f}\bra{\psi_f} \otimes \mathbbm{1}_M) \hat{\rho}_{\text{dec}} \right].$
Substituting the evolved state~(\ref{eq:rho_dec}) into the trace yields the normalized apparatus state conditional on the successful post-selection of the system:
\begin{equation}
    \hat{\rho}_M^{\text{ps}} = \sum_j \hat{\mathcal{K}}_j \hat{\rho}_{M} \hat{\mathcal{K}}_j^\dagger.
\end{equation}
Here, $\hat{\mathcal{K}}_j$ are the effective Kraus operators acting solely on the apparatus Hilbert space. To ensure that the final state, $\hat{\rho}_M^{\text{ps}}$, has unit trace, these operators $\hat{\mathcal{K}}_{j}=\langle\psi_{f}|\hat{K}_{j}\hat{U}|\psi_{i}\rangle/(P_{\text{succ}})^{1/2}$ are appropriately defined, by including the post-selection success probability $P_{\text{succ}} = \text{Tr}[\hat{\rho}^{\text{ps}}]$.
By substituting the decomposition of the unitary interaction $\hat{U}$~[Eq.(\ref{eq:Turek2015})], these effective operators can be expressed directly in terms of the displacement operators
\begin{equation}
    \hat{\mathcal{K}}_j = c_{j}^+ \hat{D}\left(\frac{s}{2}\right) + c_{j}^- \hat{D}\left(-\frac{s}{2}\right),
\end{equation}
where the normalized complex coefficients are given by
\begin{equation}\label{eq:c_j}
    c_{j}^{\pm} = \frac{1}{2\sqrt{P_{\text{succ}}}} \bra{\psi_f} \hat{K}_j ( \hat{\mathbbm{1}}_S \pm \hat{A}) \ket{\psi_i}.
\end{equation}
This decomposition reveals that, under general open dynamics, the final apparatus state is an incoherent mixture of distinct decoherence pathways~\cite{nielsen2010quantum,Petruccione_book}. By summing over these pathways and grouping the contributions associated with a common pointer displacement (see Appendix~\ref{appendix:state_simplification} for the precise definition of the shifted component states $\hat{\rho}_{\lambda\nu}$ for $\lambda,\nu \in \{+,-\}$), the total conditioned apparatus state can be expressed as
\begin{gather}
  \hat{\rho}_M^{\text{ps}} = \mathcal{W}_{++} \hat{\rho}_{++} + \mathcal{W}_{--} \hat{\rho}_{--} +  e^{-ibs}\mathcal{W}_{+-} \hat{\rho}_{+-} \nonumber \\ + e^{ibs} \mathcal{W}_{-+} \hat{\rho}_{-+}. \label{eq_rho_ps_main}  
\end{gather}
The statistical weights governing this mixture are determined by the complex coefficients $c_j^{\pm}$
\begin{gather}
\mathcal{W}_{++} = \sum_j |c_j^+|^2,\;\;\;\;\;  \mathcal{W}_{--} = \sum_j |c_j^-|^2,\;\;\;\;\;  \\
\mathcal{W}_{+-} = \sum_j c_j^+ (c_j^-)^*,\;\;\;\;  \mathcal{W}_{-+} = \mathcal{W}_{+-}^*.
\end{gather}
The real coefficients $\mathcal{W}_{++}$ and $\mathcal{W}_{--}$ represent the total statistical population of the probe shifted by $+s/2$ and $-s/2$ respectively. In contrast, the complex term $\mathcal{W}_{+-}$ encapsulates the quantum interference, which is explicitly modulated by the initial momentum of the pointer $b = \mathrm{Im}(\alpha)$ via the phase factor $e^{-ibs}$. 

\section{Thermal instabilities of weak-to-strong measurement transition}\label{sec:results}

In this section, we apply the Generalized Amplitude Damping (GAD) channel to our framework (see details in Appendix~\ref{app:GAD_model}). In doing so, we investigate how thermal instabilities affect weak value measurements. Furthermore, we explore how these thermal effects influence the success probability of correctly post-selecting the system in the desired state within the weak-value protocol. Additionally, we study the weak-to-strong measurement transition under three thermodynamic conditions, revealing the emergence of a non-monotonic transition for non-classical pointers that is entirely absent in classical-like thermal states~\cite{Pan2020Nature}. Finally, we examine the effect of the thermal bath on the measurement apparatus's sensitivity to weak-value amplification.

\subsection{Weak Value and Thermal Effects}\label{sec:weak_value}

The standard weak value, conventionally defined as $\langle A \rangle_w = \langle \psi_f | \hat{A} | \psi_i \rangle / \langle \psi_f | \psi_i \rangle$, is based on the assumption that the system evolves unitarily between the pure pre-selected and post-selected states~\cite{AharonovAlbertVaidman1988PRL}. However, for open quantum systems subject to non-unitary decoherence, this definition must be generalized to take into account the loss of coherence and the mixing of states~\cite{Wiseman2002PRA,Shikano_2010}. A rigorous generalization for pre- and post-selected mixed states is provided as~\cite{VaidmanPRA2017}
\begin{equation}
\label{eq:generalized_WV_Vaidman}
\langle A \rangle_w = \frac{\text{Tr}\left[ \hat{\rho}_{\text{post}} \hat{A} \hat{\rho}_{\text{pre}} \right]}{\text{Tr}\left[ \hat{\rho}_{\text{post}} \hat{\rho}_{\text{pre}} \right]},
\end{equation}
where $\hat{\rho}_{\text{pre}}$ and $\hat{\rho}_{\text{post}}$ represent the forward-evolving and backward-evolving states exactly at the time of the weak measurement, respectively~\cite{ABL_two_state_formalism1964,Aharonov2008}. In our model, the weak measurement interaction occurs immediately after the pre-selection, meaning the forward-evolving state is simply the pure initial state, $\hat{\rho}_{\text{pre}} = |\psi_i\rangle\langle\psi_i|$. Following the measurement, the system interacts with the thermal bath, undergoing decoherence governed by the quantum channel $\mathcal{E}_S$, before finally being post-selected onto the pure state $|\psi_f\rangle$. Consequently, to determine the effective backward-evolving state at the time of the measurement, the post-selection must be retrodicted through the thermal bath using the adjoint channel, yielding $\hat{\rho}_{\text{post}} = \mathcal{E}_S^\dagger(|\psi_f\rangle\langle\psi_f|) = \sum_k \hat{K}_k^\dagger |\psi_f\rangle\langle\psi_f| \hat{K}_k$. 
Substituting these states into Eq.~\eqref{eq:generalized_WV_Vaidman} and applying the cyclic property of the trace, the generalized weak value under this operational sequence evaluates to
\begin{equation}
\label{eq:generalized_WV}
\langle A \rangle_w^{\mathcal{E}_S} = \frac{\langle\psi_f| \mathcal{E}_S\big(\hat{A} |\psi_i\rangle\langle\psi_i|\big) |\psi_f\rangle}{\langle\psi_f| \mathcal{E}_S\big(|\psi_i\rangle\langle\psi_i|\big) |\psi_f\rangle}.
\end{equation}
This formulation encapsulates the entire effective action of the thermal bath, encompassing the weak interaction and the final post-selection. When the channel $\mathcal{E}_S$  reduces to the identity map (indicating that there is no decoherence), the trace operations simplify directly to the standard quantum mechanical inner products $\langle \psi_f | \hat{A} | \psi_i \rangle / \langle \psi_f | \psi_i \rangle$, thus recovering the original Aharonov-Albert-Vaidman weak value~\cite{AharonovAlbertVaidman1988PRL}.
\begin{figure*}[htb]
\centering
\includegraphics[scale=0.27]{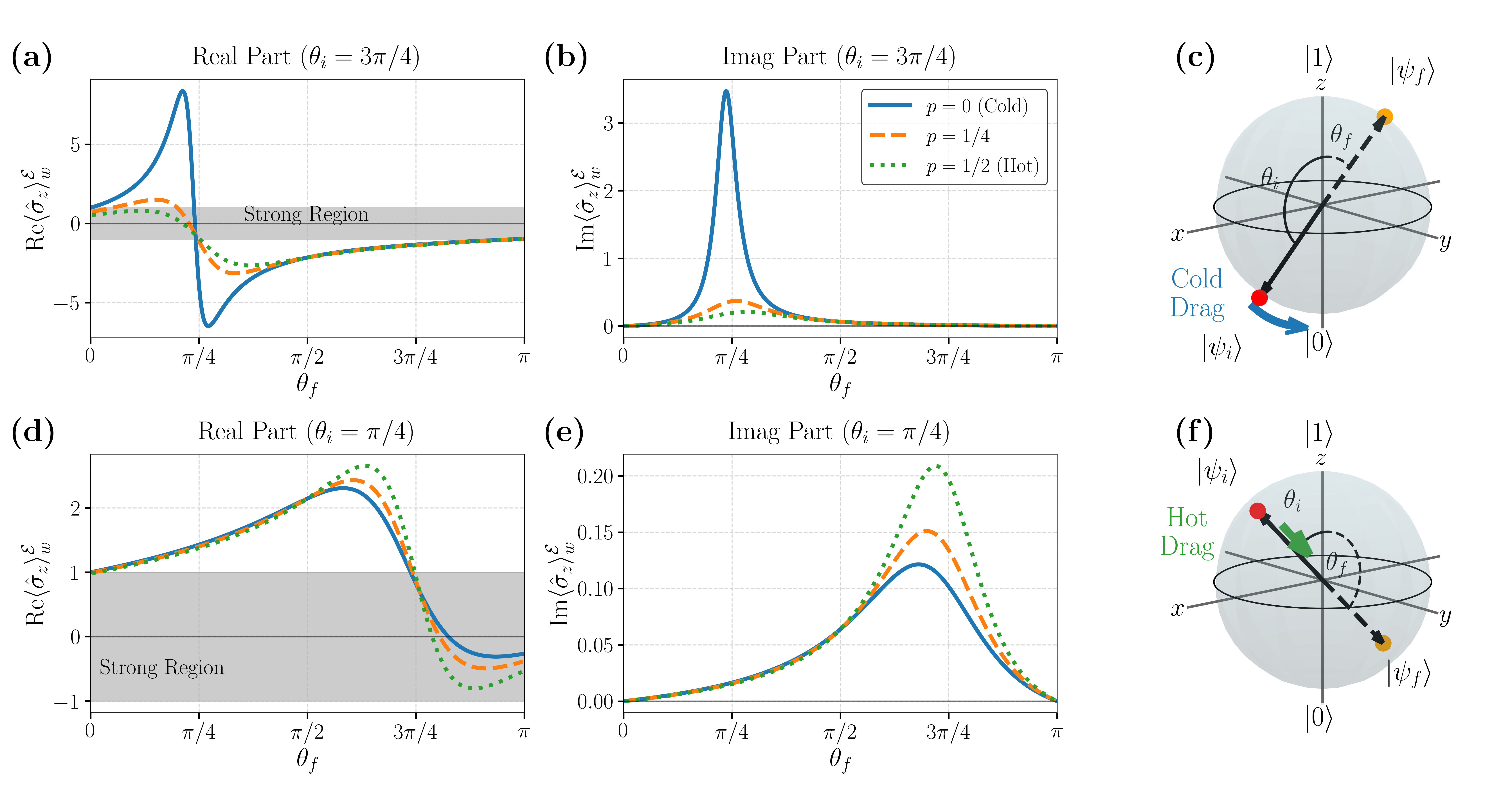}
\caption{Weak value of $\hat{\sigma}_z$ under the GAD channel as a function of the post-selection angle $\theta_f$, evaluated for a fixed damping rate $\gamma=0.1$, $\phi_i=0$, and $\phi_f=0.99\pi$. The left and middle columns display the real and imaginary parts, respectively. The gray bands in the real plots indicate the standard eigenvalue range $[-1, 1]$, beyond which anomalous amplification is achieved. Anomalies require near-orthogonal pre- and post-selected states ($\theta_f \approx \pi - \theta_i$). In the upper panels, for an initial state in the Southern hemisphere ($\theta_i = 3\pi/4$), the anomaly is maximized by a zero-temperature cold bath ($p=0$, solid blue line) and suppressed by a high-temperature hot bath ($p=1/2$, dotted green line). As shown in the Bloch sphere (c), the cold bath causes the state to decay toward $|0\rangle$ (the South pole), preserving the near-orthogonality with the Northern post-selection. In the bottom panels (d)-(f), for an initial state in the Northern hemisphere ($\theta_i = \pi/4$), the environmental role is inverted: thermal excitations from a hot bath strictly enhance the anomaly, while the cold bath suppresses it. As illustrated in (f), the hot bath contracts the state toward the origin, mitigating the decay toward $|0\rangle$. This prevents premature alignment with the required Southern post-selected state, thereby sustaining the anomaly conditions that the cold bath would otherwise destroy.}
\label{fig_weak}
\end{figure*}

In our framework, we consider the Generalized Amplitude Damping (GAD) channel $\mathcal{E}_S(\gamma,p)$ to investigate the role of thermal instabilities in the weak measurement scenario and the Pauli-$z$ as the system observable, i.e., $\hat{A} = \hat{\sigma}_z$. We emphasize that fixing the measurement operator along the $z$-axis serves merely as a choice of computational basis, not a physical restriction. To maintain complete geometric generality, we further assume that both the initial (pre-selected) state, $\ket{\psi_i} = \cos(\theta_i/2)\ket{1} + e^{i\phi_i}\sin(\theta_i/2)\ket{0}$, and the final (post-selected) state $\ket{\psi_f} = \cos(\theta_f/2)\ket{1} + e^{i\phi_f}\sin(\theta_f/2)\ket{0}$, are prepared as general superposition states. These states are parametrized, respectively, by the polar angles $\theta_{i,f} \in [0, \pi]$ and the azimuthal angles $\phi_{i,f} \in [0, 2\pi]$ on the Bloch sphere; consequently, sweeping these angles fully explores the complete SU(2) rotational parameter space of the two-level system. In turn, the corresponding weak value, defined in Eq. (\ref{eq:generalized_WV}), takes the form (see details in Appendix~\ref{appendix:sigma_z_derivation})
\begin{widetext}
\begin{gather}\label{eq:weak_z}
\langle \hat{\sigma}_z \rangle_w^{\mathcal{E}} = \frac{\cos\theta_i + \cos\theta_f \big[ 1 - \gamma + \gamma(2p-1) \cos\theta_i \big] - i\sin(\phi_i - \phi_f)\sin\theta_i\sin\theta_f\sqrt{1-\gamma}}{1 + \cos\theta_i\cos\theta_f(1-\gamma) + \gamma(2p-1)\cos\theta_f + \cos(\phi_i - \phi_f)\sin\theta_i\sin\theta_f\sqrt{1-\gamma}},
\end{gather}
\end{widetext}
where the parameters $\gamma \in [0,1]$ and $p \in [0,1/2]$ represent the strength of energy exchange with the reservoir and the thermal population of the excited state, respectively. These features of the GAD channel allow us to investigate how the combined effects of dissipation (energy exchange) and decoherence (loss of phase information in a given basis) reshape the weak value. In particular, the terms proportional to $\sin\theta_i \sin\theta_f$ correspond to the $l_1$-norm of coherence~\cite{Plenio_PRL14} of the pre- and post-selected states, which appear explicitly in both the interference contribution to the denominator and in the imaginary part of the weak value. This shows that the emergence of phase-dependent effects, especially those encoded in $\sin(\phi_i - \phi_f)$, is directly connected to the presence of quantum coherence. Moreover, these contributions are modulated by the factor $\sqrt{1-\gamma}$, indicating that the GAD channel progressively suppresses coherence and, consequently, diminishes the imaginary component of the weak value and the associated interference phenomena.

Figure~\ref{fig_weak} illustrates the impact of distinct thermodynamic regimes on the generalized weak value $\langle \hat{\sigma}_z \rangle_w^{\mathcal{E}}$. We exhibit both the real and imaginary components as a function of the post-selection angle $\theta_f$, highlighting the anomalous regions where the real part drastically exceeds the standard eigenvalue bounds $[-1, 1]$. Because anomalous amplification strictly requires near-orthogonal pre- and post-selected states ($\theta_f \approx \pi - \theta_i$), the environment's effect is highly state-dependent. For an initial state in the Southern hemisphere (upper panels), a zero-temperature cold bath ($p=0$) maximizes the anomaly. Geometrically, this occurs because the cold bath drives the state toward the South pole ($|0\rangle$), naturally preserving its near-orthogonality with the required Northern post-selection. Conversely, for a Northern initial state (lower panels), this environmental role is remarkably inverted. In this configuration, the cold bath induces a decay toward $|0\rangle$ that almost aligns the state with the required Southern post-selection, thereby destroying the necessary orthogonality and suppressing the anomaly. To counteract this, thermal excitations from a hot bath ($p=1/2$) become highly beneficial; the hot bath contracts the state toward the origin of the Bloch sphere, mitigating the downward decay and actively sustaining the geometric conditions required for anomalous amplification. It is crucial to note that this mechanism is intrinsic to the GAD channel, where the equilibrium population distribution is directly modulated by the bath temperature~\cite{nielsen2010quantum}. Furthermore, this thermal enhancement of the anomaly is not exclusive to $\hat{\sigma}_z$. Evaluating an orthogonal observable like $\hat{\sigma}_x$ yields similar dynamics, where the same counter-intuitive thermal boost fully persists, merely shifted to a rotated set of pre- and post-selection angles, confirming the broad applicability of our framework. In Sec.~\ref{sec:Amplification}, we will explore the connection between the real and imaginary parts and their significance through their effect on the measurement apparatus.

\subsection{Post-Selection Success Probability}
\label{sec:psucc_main}

\begin{figure*}[htbp]
\centering
\includegraphics[scale=0.3]{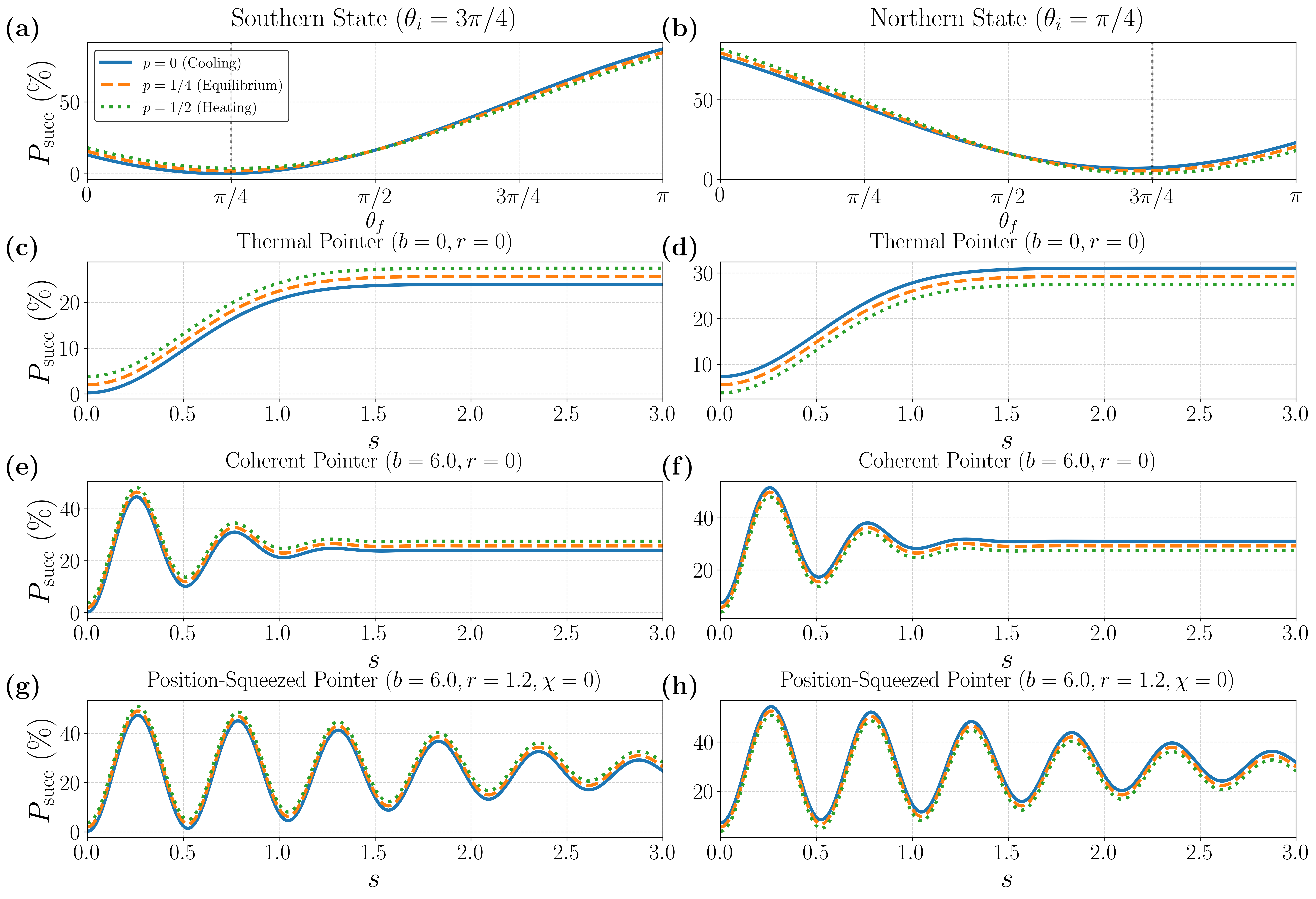}
\caption{Success probability $P_{\text{succ}}$ across distinct thermodynamic regimes for Southern ($\theta_i=3\pi/4$, left column) and Northern ($\theta_i=\pi/4$, right column) initial states. Curves compare the cooling limit ($p=0$, solid blue lines), thermal equilibrium ($p=1/4$, dashed orange lines), and heating limit ($p=1/2$, dotted green lines). (a)-(b) Angular dependence in the weak measurement regime ($s=0.01$). Vertical dotted lines mark the near-orthogonal anomalous angles ($\theta_f = \pi/4$ and $3\pi/4$). The cold bath drives $P_{\text{succ}} \to 0$ in (a), while the hot bath achieves this in (b), geometrically establishing the required conditions for anomalous weak-value amplification. (c)-(d) Measurement strength ($s$) dependence at fixed anomalous angles for an unshifted thermal pointer ($b=0, r=0$). (e)-(f) Injecting initial pointer momentum ($b=6.0$) generates dynamic quantum interference fringes. This coherence extends the protocol beyond the traditional weak limit, actively tuning $P_{\text{succ}}$ to the small, strictly non-null values necessary for practical anomalous amplification. (g)-(h) Applying position squeezing ($r=1.2, \chi=0$) to the coherent pointer drives $P_{\text{succ}}$ to even deeper minimums at periodic values of $s$. This ensures that optimal amplification conditions recur periodically even at intermediate interaction strengths, maintaining the protocol's practical viability. Fixed parameters: $\gamma=0.1$, $\bar{n}=0.5$ (yielding $p_{\text{eq}} = 1/4$), $\phi_i=0$, and $\phi_f=0.99\pi$.}
\label{fig_Psucc}
\end{figure*}

Within the framework of post-selected measurements, the success probability $P_{\text{succ}}$ represents a fundamental operational quantity. Mathematically, it provides the required normalization for the apparatus's final state, whereas physically, it dictates the statistical yield and the overall efficiency of the protocol~\cite{WalmsleyPRL2015}.  Specifically, it quantifies the likelihood of successfully projecting the system onto $\ket{\psi_f}$ in the presence of environmental decoherence. In quantum metrology, a fundamental trade-off governs this process: achieving the massive anomalous amplification necessary to enhance precision strictly requires the pre- and post-selected states to be near-orthogonal. This geometric requirement inherently enforces a highly suppressed success probability~\cite{TanakaPRA2013,KneePRA2013,KneePRX2014,FerriePRL2014}. However, the distinct advantage of the weak-value protocol lies in the fact that, although successful measurement events are rare, the drastic amplification of the pointer shift per successful event can compensate for the discarded data, ultimately benefiting the signal-to-noise ratio~\cite{Kwiat2008Science,StarlingPRA2010,SteinbergPRL2011,Wu2011PRA,JordanPRX2014}. Although evaluating this probability for open quantum systems can be analytically demanding, the specific case of the GAD channel, with the system's observable $\hat{A}=\hat{\sigma}_z$ and a displaced squeezed thermal probe, yields an exact closed-form solution (See details in Appendix~\ref{app:psucc_derivation})
\begin{gather}
 P_{\text{succ}} = \frac{1}{2} \Big[ 1 + \gamma(2p-1)\cos\theta_f + (1-\gamma)\cos\theta_i\cos\theta_f \Big]  \nonumber\\
    + \frac{e^{-\Gamma s^2}}{2} \sqrt{1-\gamma} \sin\theta_i \sin\theta_f \cos(2bs + \phi_i - \phi_f) , \label{eq:Psucc_simplified}
\end{gather}
where the squeezing and thermal parameters of the probe determine the rate $\Gamma$:
\begin{equation}
    \Gamma = (2\bar{n}+1) (\cosh 2r - \cos 2\chi \sinh 2r).
\end{equation}
Here, the imaginary part of the displacement, $b = \text{Im}(\alpha)$, sets the initial momentum of the probe. Note that Eq.~(\ref{eq:Psucc_simplified}) clearly separates the dynamics into a non-decaying background term and a quantum interference term, whose visibility is exponentially suppressed by the Gaussian decay factor $e^{-\Gamma s^2}$. Therefore, only in the limit of a weak measurement ($s \to 0$) and complete isolation from the thermal bath ($\gamma = p= 0$) does this probability naturally reduce to the standard pure-state overlap $P_{\text{succ}}= |\langle \psi_f | \psi_i \rangle|^2$ \cite{PangPRL2014}.

To analyze how the post-selection success probability is governed by thermodynamic energy exchange, we compare the meter's initial temperature $T_M$ (determined by its mean occupation number $\bar{n}$) with the bath's effective temperature $T_B$ (governed by the population parameter $p$). Assuming the meter and reservoir are resonant ($\omega_M = \omega_S$), thermal equilibrium ($T_B = T_M$) requires the reservoir's mean photon number $\bar{q}$ to equal the pointer's thermal occupation $\bar{n}$. Since the bath population parameter is defined as $p = \bar{q}/(2\bar{q}+1)$, this exact balance yields the critical threshold $p_{\text{eq}} = \bar{n}/(2\bar{n}+1)$. This value demarcates the thermodynamic behavior of the success probability $P_{\text{succ}}$: for $p < p_{\text{eq}}$ (where $T_B < T_M$), the channel acts as a heat sink, extracting energy from the apparatus (cooling regime); conversely, for $p > p_{\text{eq}}$ (where $T_B > T_M$), the environment is hotter than the pointer, injecting thermal noise into the apparatus (heating regime).

Figure~\ref{fig_Psucc} illustrates the behavior of $P_{\text{succ}}$ across these thermodynamic regimes. In the weak measurement limit (upper panels), the necessary suppression of $P_{\text{succ}}$ at near-orthogonal anomalous angles is heavily environment-dependent: a zero-temperature cold bath ($p=0$) drives $P_{\text{succ}} \to 0$ for a Southern initial state ($\theta_i=3\pi/4$), whereas a hot bath ($p=1/2$) achieves this critical geometric suppression for a Northern initial state ($\theta_i=\pi/4$). Moving beyond the traditional weak limit ($s \to 0$), the measurement-strength ($s$) dependence reveals additional operational control. Unlike the unmodulated success probability of an unshifted thermal pointer (c-d), injecting initial pointer momentum (e-f) introduces dynamic quantum interference fringes. This injected coherence provides a crucial tuning mechanism, actively driving $P_{\text{succ}}$ to the small, strictly non-null values required to sustain the anomaly at intermediate interaction strengths. Finally, applying position squeezing (g-h) deepens these minimums, ensuring that optimal amplification conditions recur periodically and maintaining the protocol's practical viability for intermediate interaction strengths.

\subsection{Weak-to-Strong Measurement Transition}
\label{sec:transition_quantification}

\begin{figure}[htb]
\centering
\includegraphics[scale = 0.55]{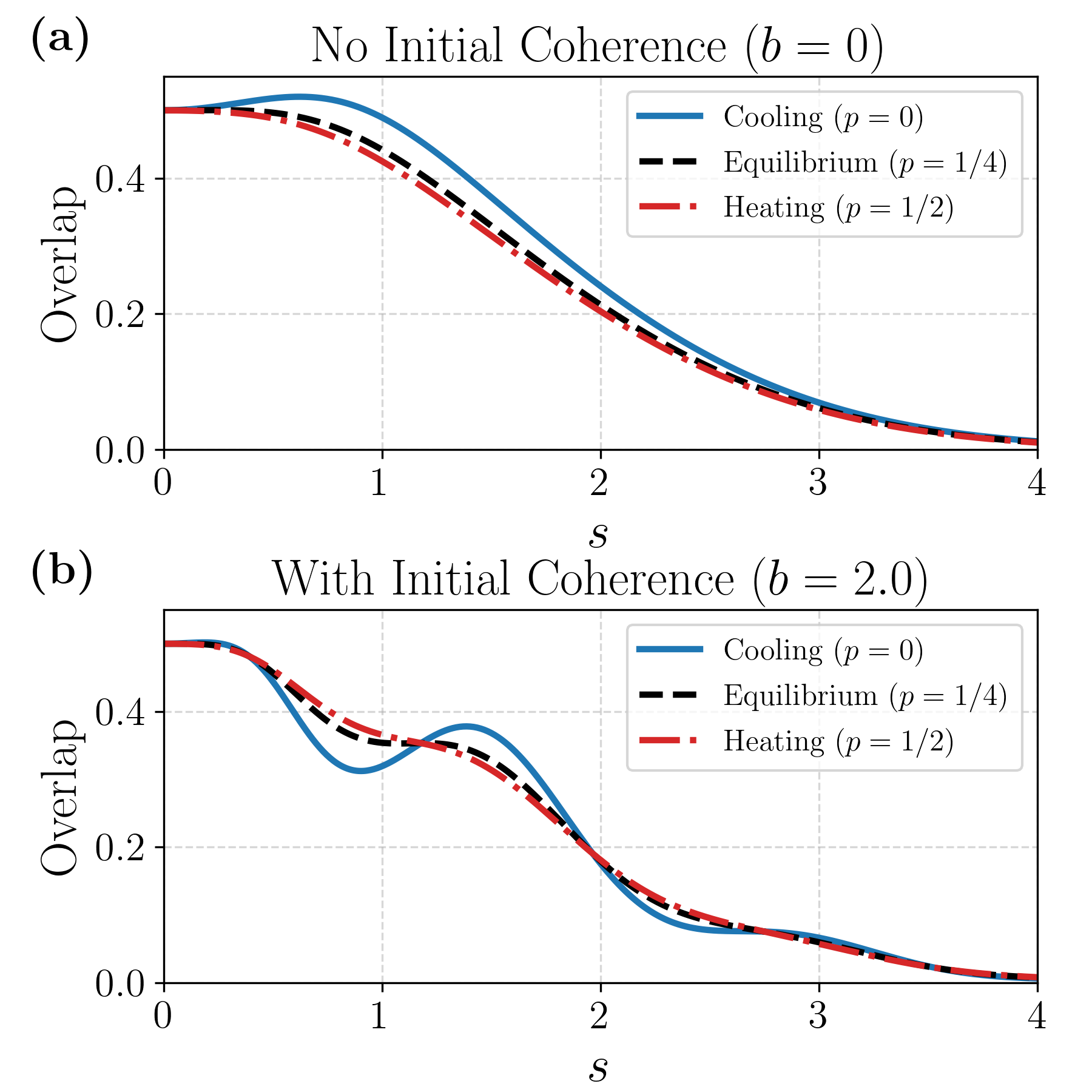}
\caption{Thermal effects in the weak-to-strong measurement transition. The quantum overlap $\mathcal{O}$ is plotted as a function of the measurement strength $s$ for three distinct thermodynamic regimes: the cooling limit ($p=0$, solid blue line), thermal equilibrium ($p=1/4$, dashed black line), and the infinite-temperature heating limit ($p=1/2$, dash-dotted red line). (a) For a purely thermal pointer with no initial momentum ($b=0$), the overlap exhibits a monotonic decay, where the heating regime strictly accelerates thermodynamic decoherence compared to the cooling limit. (b) By injecting initial coherence into the pointer ($b=2.0$), quantum interference fringes emerge during the transition. The high-temperature bath rapidly washes out these revivals in the overlap, demonstrating how a hot environment dynamically suppresses this non-monotonic complexity and accelerates the transition into the featureless, classical strong measurement regime. In both panels, the apparatus temperature is fixed at $\bar{n}=1/2$ (establishing $p_{\text{eq}} = 1/4$), with strong damping $\gamma=0.8$, zero squeezing ($\zeta=0$), and state geometry $\theta_i=\pi/4, \theta_f=\pi/8$ (with $\phi_i=\phi_f=0$).}
\label{fig_overlap_thermal}
\end{figure}

\begin{figure}[htb]
\centering
\includegraphics[scale=0.55]{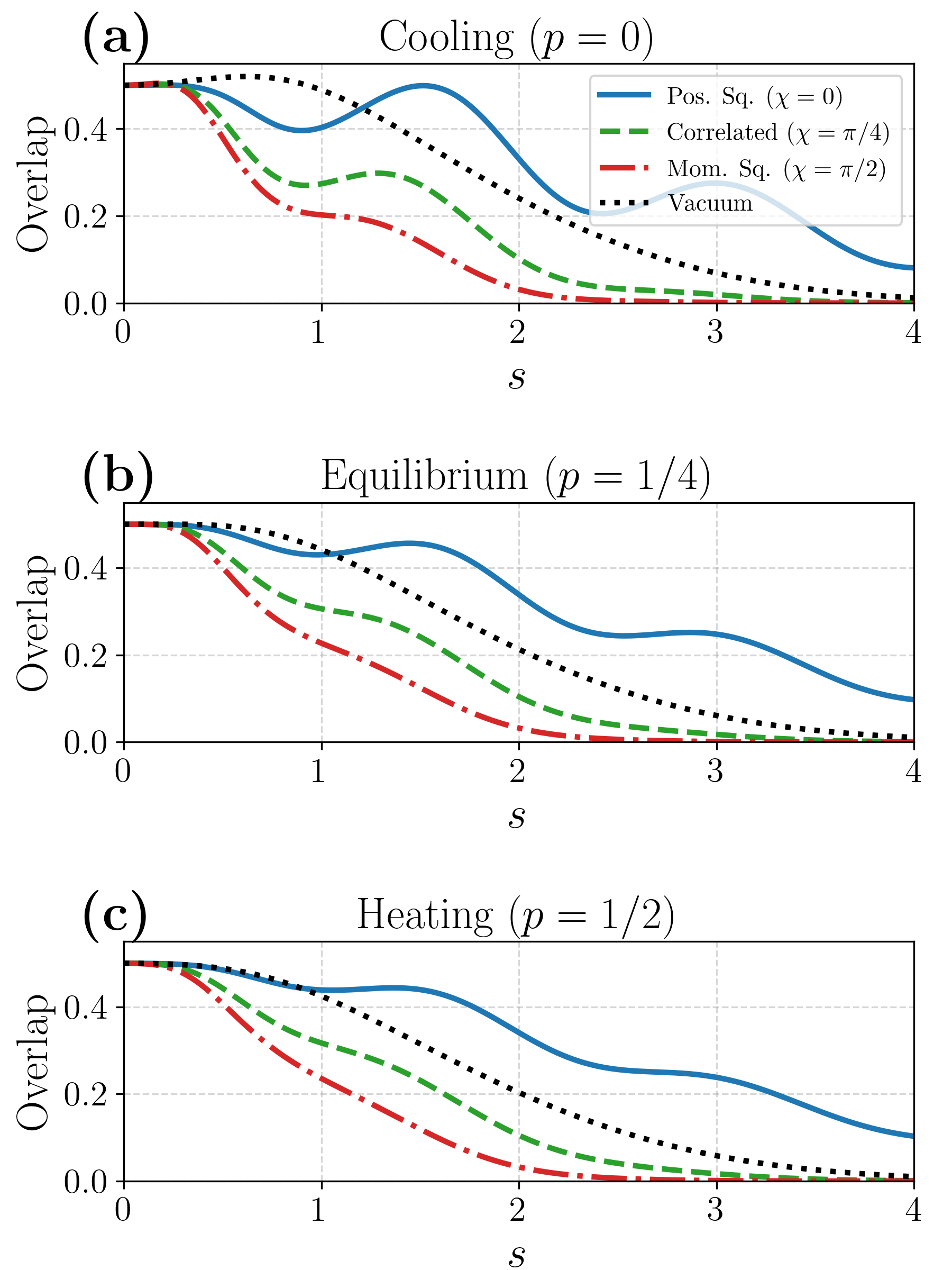}
\caption{Squeezing-phase effect in the weak-to-strong measurement transition. The quantum overlap $\mathcal{O}$ is evaluated across the three thermodynamic limits: (a) Cooling ($p=0$), (b) Thermal equilibrium ($p=1/4$), and (c) Heating ($p=1/2$). Rather than protecting the pointer, the interaction dictates that the squeezing phase $\chi$ determines whether environmental decoherence is mitigated or amplified. For a fixed squeezing magnitude ($r=0.5$) and initial momentum ($b=2.0$), aligning the squeezing along the canonical position quadrature ($\chi=0$, solid blue lines) effectively shields the pointer, sustaining revivals in the quantum overlap deeper into the strong measurement regime. These extended oscillations highlight the complex, non-monotonic nature of the transition. Conversely, squeezing along the momentum quadrature ($\chi=\pi/2$, dash-dotted red lines) amplifies the effective decay rate $\Gamma$, washing out interference fringes and forcing a rapid, monotonic decay. A correlated probe ($\chi=\pi/4$, dashed green lines) yields intermediate decay behavior. The dotted black line represents the vacuum reference probe. Other parameters are fixed to $\bar{n}=1/2$, $\gamma=0.8$, with angles $\theta_i=\pi/4$, $\theta_f=\pi/8$, and $\phi_i=\phi_f=0$.}
\label{fig_overlap_squeezing}
\end{figure}

To quantify the transition from weak to strong measurement, we evaluate the quantum overlap $\mathcal{O} = \mathrm{Tr}(\hat{\rho}_{M} \hat{\rho}_M^{\mathrm{ps}})$ between the initial apparatus state $\hat{\rho}_M$~\eqref{eq:rho_M_initial} and its final reduced state $\hat{\rho}_M^{\mathrm{ps}}$~\eqref{eq_rho_ps_main}, which is conditioned on the successful post-selection of the system. Physically, this quantity captures the operational survival of pointer indistinguishability. As established in the continuous measurement transition formalism~\cite{Pan2020Nature}, anomalous weak-value amplification is fundamentally an interference phenomenon~\cite{Berry_2012JPA} that survives only when the measuring apparatus remains largely unperturbed. By evaluating the overlap with the initial state, $\mathcal{O}(s)$  quantifies the macroscopic dislocation of the apparatus. As the interaction strength scales into the strong projective regime, the pointer wavepackets macroscopically shift away from their original unshifted position; consequently, this overlap asymptotically vanishes, signaling the loss of interference and the transition to the classical conditional expectation value. The resulting simplified analytical expression is presented below, with the full mathematical derivation detailed in Appendix~\ref{appendix:state_simplification} and Appendix~\ref{app:overlap_explicit}:
\begin{equation}
\label{eq:overlap_simplified_main}
\mathcal{O}(s) = \frac{ e^{-\frac{\widetilde{\Gamma} s^2}{2}} \left[ \eta + \Psi \right]}
{(2\bar{n}+1)(\eta + e^{-\Gamma s^2} \Psi)},
\end{equation}
where
\begin{gather}
    \widetilde{\Gamma}=\frac{1}{2\bar n+1}
\left(
\cosh 2r - \cos 2\chi\,\sinh 2r
\right),  \\    \eta = 1 + \gamma(2p-1)\cos\theta_f + (1-\gamma)\cos\theta_i\cos\theta_f, \\
         \Psi = \sqrt{1-\gamma}\sin\theta_i\sin\theta_f\cos(2bs + \phi_i - \phi_f).
\end{gather}
This analytical result generalizes the Gaussian transition factor observed in unitary trapped-ion experiments~\cite{Pan2020Nature} to the thermodynamic regime, explicitly quantifying how thermal effects modify the exponential suppression of pointer interference. Equation~\eqref{eq:overlap_simplified_main} extends this fundamental concept to open quantum systems and non-classical probes. Here, the transition is no longer dictated by a single parameter, but rather by the competition between two distinct decay scales: the global envelope rate $\widetilde{\Gamma}$, which determines the overall width of the wavepacket overlap, and the interference suppression rate $\Gamma$ (appearing in the denominator), which specifically dampens the quantum interference terms $\Psi$. Their interplay generates a non-monotonic transition behavior that is actively modulated by the initial pointer momentum $b = \mathrm{Im}(\alpha)$ and the thermodynamic competition between the bath temperature parameter $p$ and the pointer thermal occupation number $\bar{n}$.

To illustrate the physical implications of these competing decay scales, Figure~\ref{fig_overlap_thermal} analyzes how probe coherence and thermal effects impact the weak-to-strong measurement transition. For a purely thermal pointer with no initial momentum ($b=0$), the overlap exhibits a strictly monotonic decay, where the heating limit accelerates thermodynamic decoherence compared to the cooling regime. However, by injecting initial coherence into the pointer ($b=2.0$), quantum interference fringes emerge during the transition. While a cold environment preserves these revivals, a high-temperature bath dynamically washes them out, actively suppressing this non-monotonic complexity and accelerating the transition into a featureless, classical strong measurement regime. Furthermore, Figure~\ref{fig_overlap_squeezing} reveals the phase-dependent shielding effect of a squeezed pointer. Aligning the initial squeezing along the canonical position quadrature ($\chi=0$) effectively shields the pointer, protecting quantum coherence and sustaining overlap revivals deeper into the strong measurement regime, even in the infinite-temperature limit. Conversely, momentum squeezing ($\chi=\pi/2$) actively amplifies the effective decay rate $\Gamma$, forcing a rapid, monotonic decay that completely washes out the interference fringes. Finally, a probe prepared with an intermediate squeezing phase ($\chi=\pi/4$) yields an intermediate decay behavior. This configuration corresponds to a state exhibiting a non-null correlation between the position and momentum quadratures~\cite{OzielPRA2026}. In the context of the weak-to-strong transition, this correlated probe effectively balances the robust protective shielding of position squeezing with the accelerated thermal decoherence of momentum squeezing, ultimately offering a highly tunable intermediate pathway toward the strong measurement limit.

\subsection{Position and Momentum Shifts, Anomalous Amplification, and Signal-to-Noise Ratio}\label{sec:Amplification}

\begin{figure*}[htbp]
\centering
\includegraphics[scale=0.33]{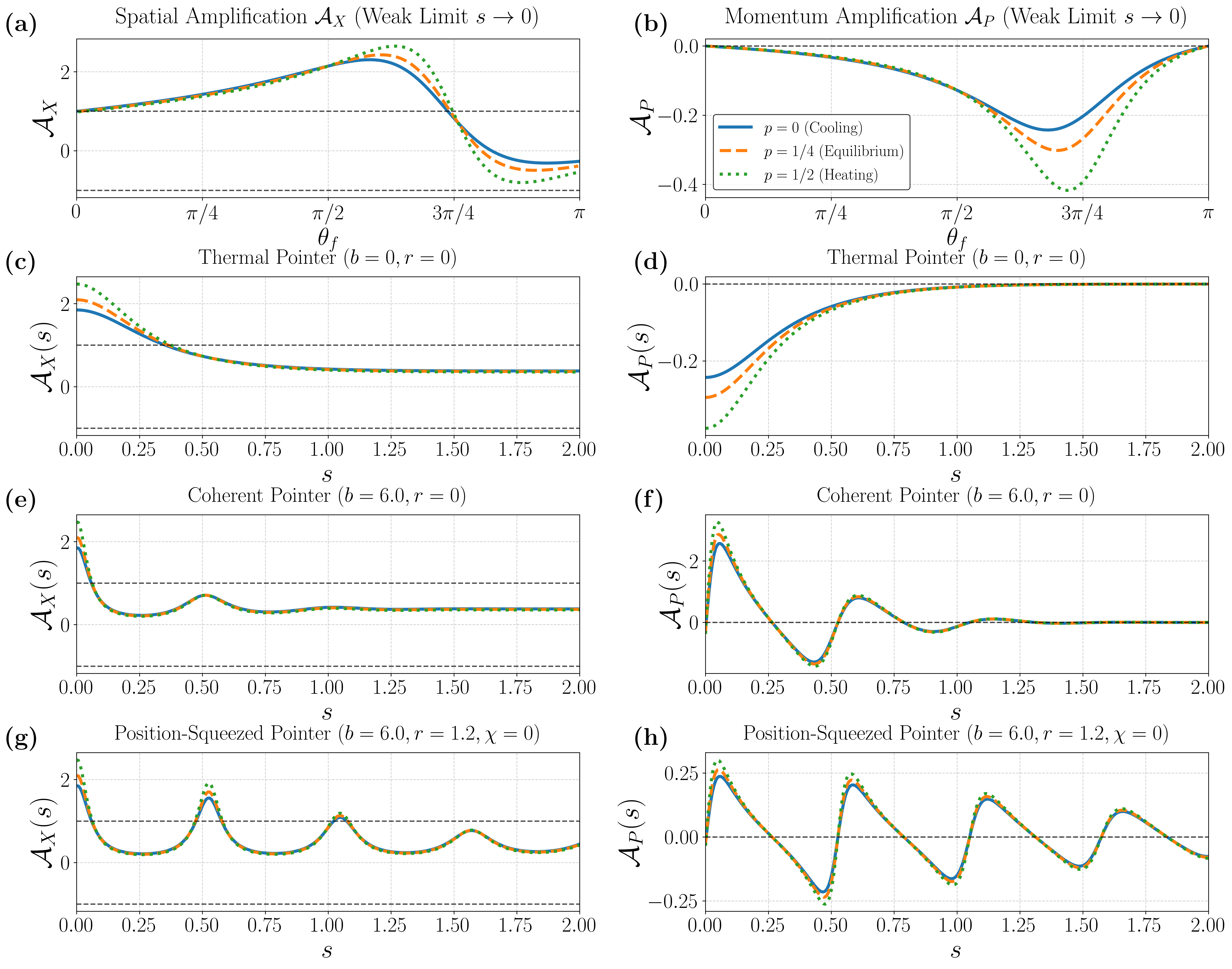}
\caption{Spatial ($\mathcal{A}_X$, left column) and momentum ($\mathcal{A}_P$, right column) amplifications across distinct thermodynamic regimes. Curves compare the cooling limit ($p=0$, solid blue lines), thermal equilibrium ($p=1/4$, dashed orange lines), and heating limit ($p=1/2$, dotted green lines). Horizontal dashed lines indicate the classical macroscopic bounds ($\mathcal{A}_X = \pm 1$ and $\mathcal{A}_P = 0$). (a)-(b) Angular dependence in the weak measurement regime ($s=0.01$). The measurement strength ($s$) dependence at the fixed anomalous angle is shown for an unshifted thermal pointer ($b=0, r=0$) in (c)-(d), and for an initial coherent momentum state ($b=6.0$) in (e)-(f). Both shifts rapidly decay to their classical limits as $s$ increases; however, the injected coherence in (e)-(f) creates intermediate regimes of $s$ where the anomaly can emerge outside the weak limit ($s \to 0$). (g)-(h) Applying position squeezing ($r=1.2, \chi=0$) to the coherent pointer further enhances the occurrence of the anomaly. Fixed parameters: $\gamma=0.1$, $\bar{n}=0.5$ (yielding $p_{\text{eq}} = 1/4$), $\phi_i=0$, and $\phi_f=0.99\pi$.}
\label{fig_amplifications}
\end{figure*}

The impact of the thermal bath on the measurement process manifests directly in the pointer dynamics; therefore, we evaluate both its net spatial and momentum shifts (with detailed derivations provided in Appendices~\ref{app:shift_X} and \ref{app:shift_P}, respectively). By summing the contributions of the conditioned pointer state $\hat{\rho}_M^{\text{ps}}$, weighted by their respective probabilities, we arrive at the final analytical results for the mean position and momentum shifts
\begin{gather}
    \Delta X = \sigma s (\mathcal{W}_{++} - \mathcal{W}_{--}), \\
    \Delta P = 2 \frac{\Gamma s}{\sigma} e^{-\Gamma s^2} \text{Im} \Big( \mathcal{W}_{+-} e^{-2isb} \Big).
\end{gather}
Note that the spatial shift is governed entirely by the diagonal (classical) elements, $\mathcal{W}_{++}$ and $\mathcal{W}_{--}$. Conversely, the off-diagonal cross-term, $\mathcal{W}_{+-}$, is responsible for the emergence of the anomalous momentum shift $\Delta P$, a purely quantum phenomenon driven exclusively by the survival of the pointer's coherence. The quantity $\sigma s$ represents the fundamental spatial shift of the pointer when perfectly correlated with a definite system eigenstate, yielding a shift of $+\sigma s$ when the system is in the $|1\rangle$ eigenstate, and $-\sigma s$ when the system is in the $|0\rangle$ eigenstate. In strict analogy, we scale the momentum shift against the canonically conjugate characteristic momentum scale of the interaction, $s/\sigma$. Consequently, the mean shifts can be factored as $\Delta X = (\sigma s) \mathcal{A}_X(s)$ and $\Delta P = (s/\sigma) \mathcal{A}_P(s)$, where we define the dimensionless amplification factors
\begin{gather}
    \mathcal{A}_X(s) \equiv \mathcal{W}_{++} - \mathcal{W}_{--}, \\
    \mathcal{A}_P(s) \equiv 2 \Gamma e^{-\Gamma s^2} \text{Im} \Big( \mathcal{W}_{+-} e^{-2isb} \Big).
\end{gather}
Substituting the explicit generalized amplitude damping (GAD) coefficients $c_j^{\pm}$ (Appendix~\ref{app:GAD-coef}), we obtain the amplification factors as explicit functions of the thermal parameters
\begin{gather} \label{eq:amplifications}
    \mathcal{A}_X(s) =  \frac{\Big\{ \cos\theta_i + \cos\theta_f \big[ 1 - \gamma + \gamma(2p-1) \cos\theta_i \big] \Big\}}{2 P_{\text{succ}}},  \\
    \mathcal{A}_P(s) =  \frac{\Big[ - \Gamma e^{-\Gamma s^2} \sqrt{1-\gamma} \sin\theta_i \sin\theta_f \sin(2bs -\Delta\phi) \Big]}{2 P_{\text{succ}}}. \label{eq:amplification_p}
\end{gather}
These results show how pre- and post-selection, together with the thermal parameters, shape the existence of anomalous weak values. Note that in the weak limit ($s \to 0$), both the spatial and momentum shifts correspond to the real and imaginary parts of the weak value~\eqref{eq:weak_z}; that is, $\lim_{s \to 0} \mathcal{A}_X(s) = \text{Re}[\langle \hat{\sigma}_z \rangle_w^{\mathcal{E}}]$ and $\lim_{s \to 0} \mathcal{A}_P(s) = \Gamma \; \text{Im}[\langle \hat{\sigma}_z \rangle_w^{\mathcal{E}}]$. This demonstrates that anomalous weak values remain linked to the pointer shifts even under thermal effects~\cite{Wu2011PRA}.  On the other hand, in the strong measurement limit ($s \to \infty$), the spatial amplification becomes restricted to the standard eigenvalue spectrum $[-1, 1]$, while the momentum amplification vanishes entirely ($\mathcal{A}_P \to 0$). This confirms that the complete loss of the pointer's coherence in this regime strictly eliminates the possibility of anomalous amplification. The explicit analytical calculation and verification demonstrating that these shifts converge to the classical conditional expectation value in the strong-measurement limit ($s \to \infty$) are detailed in the Appendix~\ref{app:shift_X}.

Figure~\ref{fig_amplifications} details the spatial ($\mathcal{A}_X$) and momentum ($\mathcal{A}_P$) amplifications across distinct thermodynamic regimes. In the weak-measurement limit (a-b), the angular dependence shows the emergence of anomalous shifts that drastically exceed their respective classical macroscopic bounds ($\mathcal{A}_X = \pm 1$ and $\mathcal{A}_P = 0$). For an unshifted thermal pointer (c-d), the anomalies rapidly vanish, collapsing into the classical bounds as $s$ increases. However, injecting initial coherent momentum (e-f) actively counteracts this rapid decay, creating intermediate interaction regimes in which the anomaly survives outside the traditional weak limit ($s \to 0$). Finally, applying position squeezing (g-h) further enhances the occurrence and robustness of these anomalies, demonstrating that tailored Gaussian pointer states can practically sustain massive amplification even for larger interaction strengths.

In order to quantify the metrological advantage of the framework, we analyze the Signal-to-Noise Ratio (SNR) of the spatial shift. Following the established benchmark for post-selected weak measurements, the SNR for an estimator based on the average measurement result over $N$ trials is proportional to the mean shift and inversely bounded by the root-variance of the pointer~\cite{PangPRL2015}:
\begin{equation}
\label{eq:SNR_exact}
\text{SNR} = \frac{\sqrt{N P_{\text{succ}}} |\Delta X|}{\sqrt{\text{Var}(\hat{X})}} = \frac{\sqrt{N P_{\text{succ}}}\; \sigma s |\mathcal{A}_X(s)|}{\sqrt{\text{Var}(\hat{X})}},
\end{equation}
where $\text{Var}(\hat{X}) = \text{Tr}(\hat{X}^2 \hat{\rho}_M^{\text{ps}}) - (\Delta X)^2$ is the variance of the pointer evaluated on the final, conditioned state. This variance can be analytically simplified to the following form (see Appendix~\ref{app:SNR}):
\begin{equation}
\label{eq:exact_variance_simplified}
\text{Var}(\hat{X}) = \sigma^2 \Gamma + \sigma^2 s^2 \left[ \frac{A \left( A + B e^{-\Gamma s^2} \right) - C^2}{\left( A + Be^{-\Gamma s^2} \right)^2} \right],
\end{equation}where we defined the following auxiliary variables:
\begin{gather}
A = 1 + \cos\theta_f \big[ (1-\gamma)\cos\theta_i + \gamma(2p-1) \big], \\
B = \sqrt{1-\gamma} \sin\theta_i \sin\theta_f \cos(2bs - \phi_f + \phi_i), \\
C  = \cos\theta_i + \cos\theta_f \big[ 1 - \gamma + \gamma(2p-1)\cos\theta_i \big].
\end{gather}

\begin{figure}[htbp]
\centering
\includegraphics[scale=0.52]{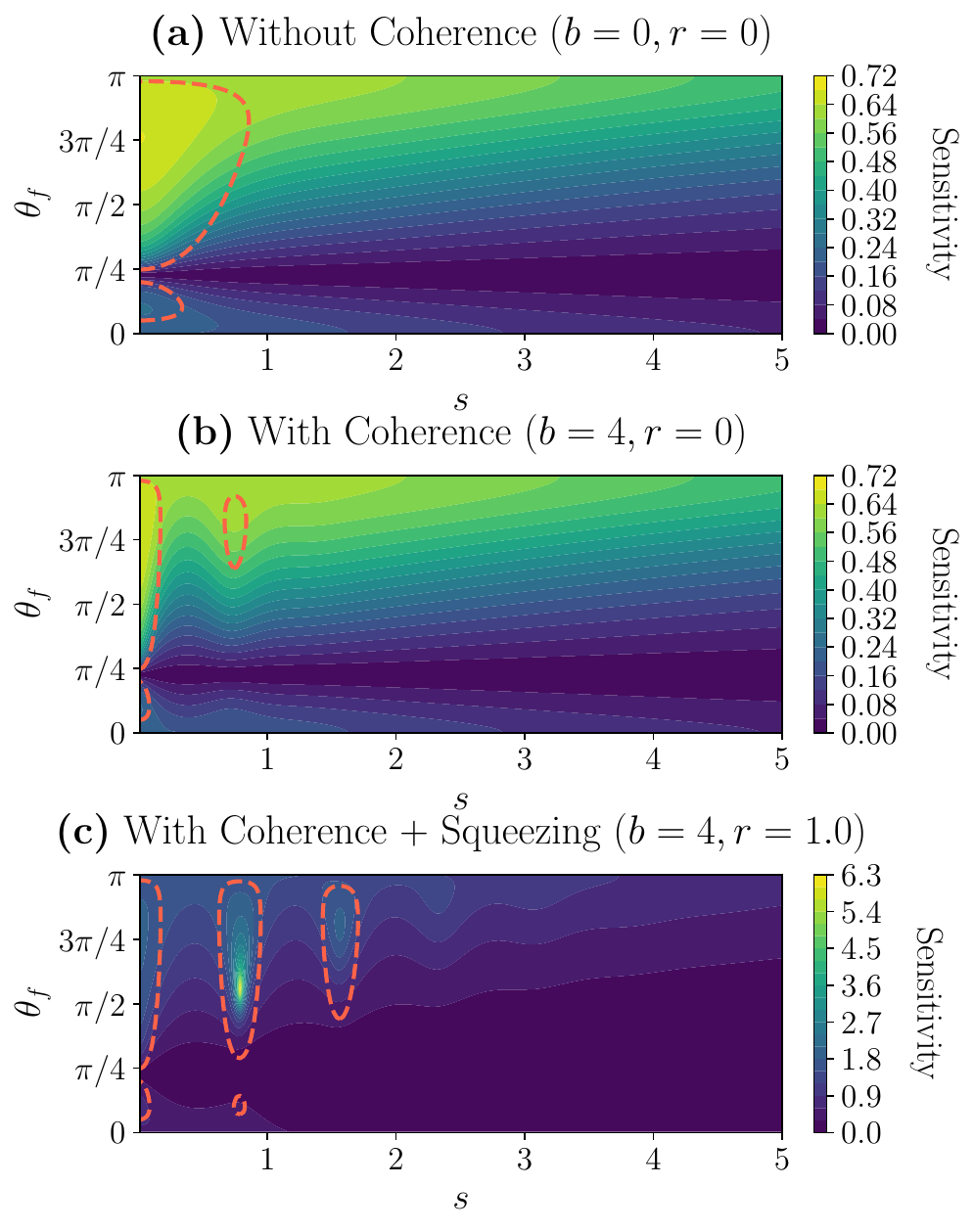}
\caption{Normalized signal-to-noise ratio, $\mathcal{S}=\text{SNR}/(s\sqrt{N})$, defined as the single-shot sensitivity,  across the measurement strength ($s$) and post-selection angle ($\theta_f$) parameter space. The dashed orange lines denote the standard classical limit ($|\mathcal{A}_X| = 1$), with the region contained within this boundary representing the anomalous regime ($|\mathcal{A}_X| > 1$). Panel (a) shows that thermal instabilities suppress the sensitivity within the anomalous regime. Panels (b) and (c) demonstrate that injecting quantum resources (specifically coherent displacement $b=4$ and squeezing $r=1.0$) restores the metrological advantage ($\mathcal{S} > 1$) and extends the optimal operational window into intermediate interaction strengths. Fixed parameters: $\bar{n}=0.5$ (yielding $p_{\text{eq}} = 1/4$), $\gamma=0.1$, $p=p_{\text{eq}} = 1/4$, $\sigma=1$, $\theta_i = 3\pi/4$, $\phi_i=0$, and $\phi_f=0.99\pi$.}  
\label{fig_SNR}
\end{figure}

To separate the linear scaling of the shift ($\Delta X \propto s$) from the protocol's inherent precision, we examine the normalized Signal-to-Noise Ratio, $\mathcal{S}=\text{SNR}/(s\sqrt{N})$, defined as the single-shot sensitivity. Figure~\ref{fig_SNR} illustrates the optimization of this sensitivity across the measurement strength ($s$) and post-selection angle ($\theta_f$). To identify regions where high precision aligns with anomaly shifts, we overlay the standard classical limit ($|\mathcal{A}_X| = 1$) represented by a dashed orange boundary; the area bounded within this contour represents the anomalous regime ($|\mathcal{A}_X| > 1$). For a thermal pointer lacking initial coherence or squeezing [Fig.~\ref{fig_SNR}(a)], even inside regions of anomalous shift, the presence of thermal instabilities strictly reduces the sensitivity below 1 ($\mathcal{S} < 1$). Thus, although an anomalous spatial shift occurs, the practical applicability of the protocol for metrological applications is severely reduced. However, the introduction of initial coherence ($b=4$) [Fig.~\ref{fig_SNR}(b)] and squeezing ($r=1.0$) [Fig.~\ref{fig_SNR}(c)] suppresses the  thermal noise, amplifying the sensitivity within the anomalous region and expanding the protocol's optimal operational window.

\section{DISCUSSION}\label{sec:discussion}

In this work, we developed a general theoretical framework to investigate the continuous weak-to-strong measurement transition under the simultaneous influence of thermal instabilities and pre- and post-selection, considering an open dynamics. By deriving the exact conditioned state of the apparatus, we demonstrated that environmental decoherence does not merely degrade the measurement signal, but rather reshapes the weak-value statistics and the measurement crossover in a highly nontrivial manner.

Applying this to the Generalized Amplitude Damping (GAD) channel revealed a counterintuitive interplay between thermal noise and the weak-value anomaly: increasing the bath temperature can actually enhance the anomalous weak value. In the vacuum limit, spontaneous emission imposes a strong directional bias that rapidly destroys the near-orthogonality required for amplification. In contrast, thermal excitations in a high-temperature bath counterbalance this decay, effectively supporting the system's excited population and preserving the geometric conditions necessary for a large anomalous signal. Regarding the measurement crossover, our results generalize the Gaussian transition factor~\cite{Pan2020Nature} to non-classical probes. By injecting initial coherence into the pointer, the transition becomes fundamentally non-monotonic, characterized by quantum revivals that extend beyond a simple decay envelope. Crucially, this transition can be actively controlled via phase-dependent squeezing. Aligning the squeezing phase with the appropriate canonical quadrature, or employing position-momentum correlated states~\cite{OzielPRA2026}, effectively shields the apparatus against environmental noise, offering a highly tunable pathway toward the strong measurement limit.

In practical experimental settings, overcoming systematic errors poses a fundamental challenge, as fixed technical limitations, such as a detector's finite resolution, cannot be mitigated through repeated statistical trials. Here, the weak-value protocol offers a distinct metrological advantage. By translating a microscopic quantum signal into a macroscopic pointer shift, anomalous amplification makes the signal large enough to be reliably read by imperfect laboratory equipment, effectively raising it above the background. This robust macroscopic signal, coupled with the extreme sensitivity of the weak value to the bath's thermal parameter $p$, enables its direct application as a highly precise quantum sensor. Because the anomalous amplification shifts continuously between the limits of a cold and a hot bath, tuning the pre- and post-selection angles to the near-orthogonal regime allows the pointer to act as a precise thermometric probe. Ultimately, these findings provide new insights into how thermal instabilities can be incorporated into the design of optimized metrology protocols, actively leveraging anomalous amplification to enhance sensing precision across the full weak-to-strong crossover.

Ultimately, our results establish that while thermal instabilities challenge the standard post-selected measurement, the strategic integration of initial coherence and squeezing provides a robust framework to harness anomalous amplification for high-precision metrological applications across the entire weak-to-strong measurement transition.

From an experimental perspective, testing this thermodynamic framework is highly feasible and offers a distinct practical advantage over idealized closed-system models. The requisite thermal instabilities can be naturally engineered in trapped-ion setups~\cite{Pan2020Nature} by coupling the ion's internal states to a controlled reservoir prior to post-selection. Alternatively, photonic platforms already employ incoherent, mixed-state pointers~\cite{Cho_2010NJP} alongside mature continuous-variable techniques like squeezing and homodyne detection~\cite{Gao_PhysRevApplied2024,MartinelliPRL2022,TabosaPRA2026,FelintoPRR2026}. Crucially, because our framework explicitly accommodates finite-temperature noise, it actively relaxes the immense technical burden of achieving zero-temperature isolation. By exploiting non-unitary thermal dynamics rather than suppressing them, exploring these non-trivial thermodynamic crossovers falls firmly within the reach of current state-of-the-art quantum technologies.

\begin{acknowledgments}M. V. S. Lima acknowledges the Federal University of Piauí (UFPI) for funding his scientific initiation scholarship. C.H.S.V. acknowledges the São Paulo Research Foundation (FAPESP) Grant No. 2023/13362-0 and Grant No. 2025/14546-2 for financial support and the Southern University of Science and Technology (SUSTech) for providing the workspace during the internship. I.G.P. acknowledges Grant No. 306528/2023-1 from CNPq. P.R.D. acknowledges support from CNPq/MCTI/FNDTC 22/2024 (No. 446775/2024-0). L.S.M. acknowledges Grant No. 3305943/2026-0 from CNPq and support from the National Institute of Science and Technology on National Institute of Photonics (INFO) CNPq - INCT grant 409174/2024-6.
\end{acknowledgments}

\vspace{0.5cm}
\noindent \textbf{Data availability statement:} All data that support the findings of this study are included within the article.

\vspace{0.2cm}
\noindent \textbf{Conflict of interest:} The authors declare that they have no conflicts of interest.

\appendix

\section{Generalized Amplitude Damping Channel}
\label{app:GAD_model}
To illustrate this framework, we consider the Generalized Amplitude Damping (GAD) channel, which models the dissipation of a qubit system interacting with a bosonic bath at finite temperature~\cite{nielsen2010quantum}. The channel is parametrized by $\gamma \in [0,1]$ and $p \in [0,1/2]$, representing the probability of a photon exchange with the reservoir and the equilibrium population of the excited state, respectively. The bath properties govern this equilibrium population according to the relation~\cite{BanerjeePRA2008}
\begin{equation}
  p = \frac{\bar{q}}{2\bar{q} + 1},
\end{equation}
where $\bar{q} = [\exp(\hbar \omega_S / k_B T_B) - 1]^{-1}$ is the mean photon number of the reservoir at the system transition frequency $\omega_S$ and temperature $T_B$. The dynamics are described by the set of Kraus operators $\{ \hat{K}_0, \hat{K}_1, \hat{K}_2, \hat{K}_3 \}$ defined in the computational basis $\{ \ket{0}, \ket{1} \}$, corresponding to the eigenstates of $\hat{\sigma}_z$~($\hat{\sigma}_z\ket{1}=+\ket{1}$, $\hat{\sigma}_z\ket{0}=-\ket{0}$)~\cite{nielsen2010quantum}
\begin{gather}
    \hat{K}_0 = \sqrt{1-p} \left( \ket{0}\bra{0} + \sqrt{1-\gamma} \ket{1}\bra{1} \right),\nonumber \\
    \hat{K}_1 = \sqrt{(1-p)\gamma} \ket{0}\bra{1},\nonumber \\
    \hat{K}_2 = \sqrt{p} \left( \sqrt{1-\gamma} \ket{0}\bra{0} + \ket{1}\bra{1} \right), \;\;
    \hat{K}_3 = \sqrt{p\gamma} \ket{1}\bra{0}.
\end{gather}
Physically, $\gamma$ dictates the strength of this dissipation, spanning from complete decoupling from the environment ($\gamma=0$) to complete relaxation ($\gamma=1$). In the zero-temperature limit ($T \to 0$), the mean photon number vanishes ($\bar{q} \to 0$), resulting in $p \to 0$. Conversely, in the high-temperature limit ($T \to \infty$), the thermal photon number diverges ($\bar{q} \to \infty$), leading to $p \to 1/2$. In this limit, the system approaches a maximally mixed state equilibrium due to the balance between thermal excitation and relaxation processes.

For the measurement interaction, we choose the Pauli-$Z$ operator as the system observable, i.e., $\hat{A} = \hat{\sigma}_z$. We consider that both the initial (pre-selected) state $\ket{\psi_i} = \cos(\theta_i/2)\ket{1} + e^{i\phi_i}\sin(\theta_i/2)\ket{0}$ and the final (post-selected) state $\ket{\psi_f} = \cos(\theta_f/2)\ket{1} + e^{i\phi_f}\sin(\theta_f/2)\ket{0}$ are prepared in general superposition states. These states are characterized, respectively, by the polar angles $\theta_{i,f} \in [0, \pi]$ and the azimuthal angles $\phi_{i,f} \in [0, 2\pi]$ on the Bloch sphere. The specific coefficients $c_j^{\pm}$ [see Eq.~\eqref{eq:c_j}] for the GAD channel are derived in the Appendix~\ref{app:GAD-coef}. This yields the following coefficients
\begin{gather}
    c_0^+ = \frac{\sqrt{(1-p)(1-\gamma)}}{\sqrt{P_{\text{succ}}}} \cos\frac{\theta_f}{2} \cos\frac{\theta_i}{2}, \nonumber \\
    c_0^- = \frac{\sqrt{1-p}}{\sqrt{P_{\text{succ}}}} e^{i(\phi_i-\phi_f)} \sin\frac{\theta_f}{2} \sin\frac{\theta_i}{2}, \nonumber  \\
    c_1^+ = \frac{\sqrt{(1-p)\gamma}}{\sqrt{P_{\text{succ}}}} e^{-i\phi_f}\sin\frac{\theta_f}{2} \cos\frac{\theta_i}{2}, \;\;
    c_1^- = 0, \nonumber  \\
    c_2^+ = \frac{\sqrt{p}}{\sqrt{P_{\text{succ}}}} \cos\frac{\theta_f}{2}\cos\frac{\theta_i}{2}, \nonumber \\
    c_2^- = \frac{\sqrt{p(1-\gamma)}}{\sqrt{P_{\text{succ}}}} e^{i(\phi_i-\phi_f)}\sin\frac{\theta_f}{2}\sin\frac{\theta_i}{2}, \nonumber  \\
    c_3^+ = 0, \;\;\;
    c_3^- = \frac{\sqrt{p\gamma}}{\sqrt{P_{\text{succ}}}} e^{i\phi_i} \cos\frac{\theta_f}{2} \sin\frac{\theta_i}{2}.
\end{gather}

The explicit determination of these coefficients allows us to evaluate the relevant quantities, as shown below.

\section{Derivation of GAD Coefficients}\label{app:GAD-coef}

In this appendix, we explicitly derive the normalized coefficients $c_j^{\pm}$ for each Kraus operator of the GAD channel, considering the observable system $\hat{A}=\hat{\sigma}_z$. The expression defines these coefficients:
\begin{equation}
    c_j^{\pm} = \frac{1}{2\sqrt{P_{\text{succ}}}} \left[ \bra{\psi_f}\hat{K}_j\ket{\psi_i} \pm \bra{\psi_f}\hat{K}_j\hat{\sigma}_z\ket{\psi_i} \right].
\end{equation}
We utilize the general initial state $\ket{\psi_i} = \cos(\theta_i/2)\ket{1} + e^{i\phi_i}\sin(\theta_i/2)\ket{0}$. Assuming the standard convention $\hat{\sigma}_z = \ket{1}\bra{1} - \ket{0}\bra{0}$, the operation of $\hat{\sigma}_z$ on the initial state yields $\hat{\sigma}_z\ket{\psi_i} = \cos(\theta_i/2)\ket{1} - e^{i\phi_i}\sin(\theta_i/2)\ket{0}$. The post-selection state is given by $\bra{\psi_f} = \cos(\theta_f/2)\bra{1} + e^{-i\phi_f}\sin(\theta_f/2)\bra{0}$.

\subsection{The \texorpdfstring{$c_0^{\pm}$}{} coefficients}
The operator $\hat{K}_0 = \sqrt{1-p} (\ket{0}\bra{0} + \sqrt{1-\gamma}\ket{1}\bra{1})$ describes the conditional evolution of the system in the absence of a decay event. First, we evaluate the term involving the initial state:
\begin{gather} \bra{\psi_f}\hat{K}_0\ket{\psi_i} =  \nonumber \\ \sqrt{1-p} \bra{\psi_f} \left( e^{i\phi_i}\sin\frac{\theta_i}{2}\ket{0} + \sqrt{1-\gamma}\cos\frac{\theta_i}{2}\ket{1} \right) 
    =  \nonumber \\ \sqrt{1-p} \left[ e^{i(\phi_i-\phi_f)}\sin\frac{\theta_f}{2}\sin\frac{\theta_i}{2} + \sqrt{1-\gamma}\cos\frac{\theta_f}{2}\cos\frac{\theta_i}{2} \right].
\end{gather}
Next, the term that involves the application of the operator $\hat{K}_0$ to $\hat{\sigma}_z\ket{\psi_i}$:
\begin{gather}
\bra{\psi_f}\hat{K}_0\hat{\sigma}_z\ket{\psi_i} =  \nonumber \\ 
\sqrt{1-p} \bra{\psi_f} \left( -e^{i\phi_i}\sin\frac{\theta_i}{2}\ket{0} + \sqrt{1-\gamma}\cos\frac{\theta_i}{2}\ket{1} \right) 
    =  \nonumber \\ \sqrt{1-p} \left[ -e^{i(\phi_i-\phi_f)}\sin\frac{\theta_f}{2}\sin\frac{\theta_i}{2} + \sqrt{1-\gamma}\cos\frac{\theta_f}{2}\cos\frac{\theta_i}{2} \right].
\end{gather}
Combining these results yields:
\begin{gather}
    c_0^+ = \frac{\sqrt{(1-p)(1-\gamma)}}{\sqrt{P_{\text{succ}}}} \cos\frac{\theta_f}{2} \cos\frac{\theta_i}{2},  \nonumber \\
    c_0^- = \frac{\sqrt{1-p}}{\sqrt{P_{\text{succ}}}} e^{i(\phi_i-\phi_f)} \sin\frac{\theta_f}{2} \sin\frac{\theta_i}{2}.
\end{gather}

\subsection{The \texorpdfstring{$c_1^{\pm}$}{} coefficients}
The operator $\hat{K}_1 = \sqrt{(1-p)\gamma}\ket{0}\bra{1}$ corresponds to the spontaneous emission event. For the first term, only the $\ket{1}$ component of $\ket{\psi_i}$ contributes:
\begin{gather}
\bra{\psi_f}\hat{K}_1\ket{\psi_i} = \sqrt{(1-p)\gamma} \cos\frac{\theta_i}{2} \braket{\psi_f}{0}  \nonumber \\=  \sqrt{(1-p)\gamma} e^{-i\phi_f}\sin\frac{\theta_f}{2}\cos\frac{\theta_i}{2}.
\end{gather}
For the second term, because $\hat{\sigma}_z$ leaves the $\ket{1}$ state with a positive sign, we obtain an identical result:
\begin{gather}
\bra{\psi_f}\hat{K}_1\hat{\sigma}_z\ket{\psi_i} = \sqrt{(1-p)\gamma} \cos\frac{\theta_i}{2} \braket{\psi_f}{0}  \nonumber \\
= \sqrt{(1-p)\gamma} e^{-i\phi_f}\sin\frac{\theta_f}{2}\cos\frac{\theta_i}{2}.
\end{gather}
Thus, the coefficients for the relaxation channel become:
\begin{subequations}
\begin{align}
    c_1^+ &= \frac{\sqrt{(1-p)\gamma}}{\sqrt{P_{\text{succ}}}} e^{-i\phi_f}\sin\frac{\theta_f}{2} \cos\frac{\theta_i}{2}, &
    c_1^- &= 0.
\end{align}
\end{subequations}

\subsection{The \texorpdfstring{$c_2^{\pm}$}{} coefficients}
The operator $\hat{K}_2 = \sqrt{p} (\sqrt{1-\gamma}\ket{0}\bra{0} + \ket{1}\bra{1})$ accounts for the non-unitary evolution due to the thermal bath when no photon exchange occurs.
The first term is:
\begin{gather}
    \bra{\psi_f}\hat{K}_2\ket{\psi_i} = 
    \sqrt{p} \bra{\psi_f} \left( \sqrt{1-\gamma}e^{i\phi_i}\sin\frac{\theta_i}{2}\ket{0} + \cos\frac{\theta_i}{2}\ket{1} \right) 
   \nonumber \\ = \sqrt{p} \Bigg[ \sqrt{1-\gamma}e^{i(\phi_i-\phi_f)}\sin\frac{\theta_f}{2}\sin\frac{\theta_i}{2} + \cos\frac{\theta_f}{2}\cos\frac{\theta_i}{2} \Bigg].
\end{gather}
The term acting on $\hat{\sigma}_z\ket{\psi_i}$ introduces a negative sign to the ground state component:
\begin{gather}
\bra{\psi_f}\hat{K}_2\hat{\sigma}_z\ket{\psi_i} =   \sqrt{p} \bra{\psi_f} \Big( -\sqrt{1-\gamma}e^{i\phi_i}\sin\frac{\theta_i}{2}\ket{0} + \cos\frac{\theta_i}{2}\ket{1} \Big) 
  \nonumber \\  = \sqrt{p} \left[ -\sqrt{1-\gamma}e^{i(\phi_i-\phi_f)}\sin\frac{\theta_f}{2}\sin\frac{\theta_i}{2} + \cos\frac{\theta_f}{2}\cos\frac{\theta_i}{2} \right].
\end{gather}
Combining these yields:
\begin{gather}
    c_2^+ = \frac{\sqrt{p}}{\sqrt{P_{\text{succ}}}} \cos\frac{\theta_f}{2}\cos\frac{\theta_i}{2}, \nonumber \\
    c_2^- = \frac{\sqrt{p(1-\gamma)}}{\sqrt{P_{\text{succ}}}} e^{i(\phi_i-\phi_f)}\sin\frac{\theta_f}{2}\sin\frac{\theta_i}{2}.
\end{gather}

\subsection{The \texorpdfstring{$c_3^{\pm}$}{} coefficients}
The operator $\hat{K}_3 = \sqrt{p\gamma}\ket{1}\bra{0}$ describes the absorption of a thermal photon. For the first term, only the $\ket{0}$ component of $\ket{\psi_i}$ contributes:
\begin{gather}
    \bra{\psi_f}\hat{K}_3\ket{\psi_i} = \sqrt{p\gamma} e^{i\phi_i}\sin\frac{\theta_i}{2} \braket{\psi_f}{1} 
    = \sqrt{p\gamma} \cos\frac{\theta_f}{2} e^{i\phi_i}\sin\frac{\theta_i}{2}.
\end{gather}
For the second term, the $\hat{\sigma}_z$ operator flips the sign of the $\ket{0}$ state, yielding:
\begin{gather}
\bra{\psi_f}\hat{K}_3\hat{\sigma}_z\ket{\psi_i} = -\sqrt{p\gamma} e^{i\phi_i}\sin\frac{\theta_i}{2} \braket{\psi_f}{1}
\nonumber \\ = -\sqrt{p\gamma} \cos\frac{\theta_f}{2} e^{i\phi_i}\sin\frac{\theta_i}{2}.
\end{gather}
Because the signs are strictly opposite, the $+$ coefficient cancels:
\begin{subequations}
\begin{align}
    c_3^+ &= 0, &
    c_3^- &= \frac{\sqrt{p\gamma}}{\sqrt{P_{\text{succ}}}} e^{i\phi_i} \cos\frac{\theta_f}{2} \sin\frac{\theta_i}{2}.
\end{align}
\end{subequations}

\section{Simplification of the Conditioned Apparatus State}
\label{appendix:state_simplification}

The final conditioned apparatus state is given by the incoherent sum over all decoherence pathways $j$:
\begin{equation}
    \hat{\rho}_M^{\mathrm{ps}} = \sum_j \hat{\rho}_j, \quad \text{with} \quad \hat{\rho}_j = \hat{\mathcal{K}}_j \hat{\rho}_{M} \hat{\mathcal{K}}_j^\dagger.
\end{equation}
Here, the effective Kraus operators $\hat{\mathcal{K}}_j$ acting on the apparatus. As derived in the main text, these operators are superpositions of displacement operators shifted by $\pm s/2$:
\begin{equation}
    \hat{\mathcal{K}}_j = c_{j}^+ \hat{D}\left(\frac{s}{2}\right) + c_{j}^- \hat{D}\left(-\frac{s}{2}\right),
\end{equation}
where the coefficients $c_j^{\pm}$ depend on the specific channel parameters and post-selection angles (see Appendix~\ref{app:GAD-coef}).

To simplify the calculation, we express the initial apparatus state as a displaced reference state, $\hat{\rho}_M = \hat{D}(\alpha)\hat{\rho}_0\hat{D}^\dagger(\alpha)$, where $\hat{\rho}_{0} \equiv \hat{S}(\zeta) \hat{\rho}_{\mathrm{th}} \hat{S}^\dagger(\zeta)$ is centered at the origin of phase space. To evaluate the state components $\hat{\rho}_j$, we combine the interaction shifts $\hat{D}(\pm s/2)$ with the initial displacement $\hat{D}(\alpha)$ using the Weyl composition rule~\cite{gerry2005introductory}
\begin{equation}
    \hat{D}\left(\pm\frac{s}{2}\right) \hat{D}(\alpha) = e^{\pm i \phi_\alpha} \hat{D}_{\pm},
\end{equation}
where $\hat{D}_{\pm} \equiv \hat{D}(\alpha \pm s/2)$. For a complex displacement $\alpha = a + ib$, this phase depends exclusively on the initial momentum $b$, explicitly given by $\phi_\alpha = \mathrm{Im}(\frac{s}{2}\alpha^*) = -bs/2$. We absorb these factors into the effective coefficients $C_{j}^{\pm}$:
\begin{equation}
    C_{j}^{\pm} = c_{j}^{\pm} e^{\pm i \phi_\alpha} = c_{j}^{\pm} e^{\mp i \frac{bs}{2}}.
\end{equation}
With these definitions, the state components become $\hat{\rho}_j = (C_j^+ \hat{D}_+ + C_j^- \hat{D}_-) \hat{\rho}_0 (C_j^+ \hat{D}_+ + C_j^- \hat{D}_-)^\dagger$. Expanding this product, we observe that the phase factors cancel in the diagonal terms ($|e^{\mp i bs/2}|^2=1$) and combine constructively in the off-diagonal terms ($e^{-ibs/2} \cdot e^{-ibs/2} = e^{-ibs}$). This yields:
\begin{gather}
    \hat{\rho}_j = |c_{j}^{+}|^2 \hat{D}_{+} \hat{\rho}_{0} \hat{D}_{+}^\dagger 
    + |c_{j}^{-}|^2 \hat{D}_{-} \hat{\rho}_{0} \hat{D}_{-}^\dagger 
    \nonumber \\ + c_{j}^{+} (c_{j}^{-})^* e^{-ibs} \hat{D}_{+} \hat{\rho}_{0} \hat{D}_{-}^\dagger 
    + (c_{j}^{+})^* c_{j}^{-} e^{ibs} \hat{D}_{-} \hat{\rho}_{0} \hat{D}_{+}^\dagger.
\end{gather}
Finally, summing the components  $\hat{\rho}_j$, we obtain the total conditioned state:
\begin{gather}
  \hat{\rho}_M^{\text{ps}} = \mathcal{W}_{++} \hat{\rho}_{++} + \mathcal{W}_{--} \hat{\rho}_{--} \nonumber \\ +  e^{-ibs}\mathcal{W}_{+-} \hat{\rho}_{+-} + e^{ibs} \mathcal{W}_{-+} \hat{\rho}_{-+},  
\end{gather}
where the component operators are defined as $\hat{\rho}_{\lambda\nu} = \hat{D}_\lambda \hat{\rho}_0 \hat{D}_\nu^\dagger$ (for $\lambda,\nu \in \{+,-\}$), and the coefficients are:
\begin{gather}
\mathcal{W}_{++} = \sum_j |c_j^+|^2,\;\;\;\;  \mathcal{W}_{--} = \sum_j |c_j^-|^2, \nonumber \\
\mathcal{W}_{+-} = \sum_j c_j^+ (c_j^-)^*,\;\;\;\;  \mathcal{W}_{-+} = \mathcal{W}_{+-}^*.
\end{gather}
The real coefficients $\mathcal{W}_{++}$ and $\mathcal{W}_{--}$ represent the total population of the probe shifted by $+s/2$ and $-s/2$ respectively, while the complex term $\mathcal{W}_{+-}$ encapsulates the quantum interference, modulated explicitly by the initial momentum $b$ via the phase factor $e^{-ibs}$.

\section{Weak Value under the GAD Channel}
\label{appendix:sigma_z_derivation}

In this appendix, we provide the derivation of the generalized weak value for the Pauli-$Z$ observable, $\hat{A} = \hat{\sigma}_z$, under the GAD channel. Following the operational definition established in the main text, the weak value is given by:
\begin{equation}
\langle \hat{\sigma}_z \rangle_w^{\mathcal{E}} = \frac{\bra{\psi_f} \mathcal{E}_S\big(\hat{\sigma}_z |\psi_i\rangle\langle\psi_i|\big) \ket{\psi_f}}{\bra{\psi_f} \mathcal{E}_S\big(|\psi_i\rangle\langle\psi_i|\big) \ket{\psi_f}}.
\end{equation}
In the following, we evaluate the denominator and the numerator separately.

\subsection{Evaluation of the Denominator}

To determine the denominator, we first calculate the action of the channel on the initial state $|\psi_i\rangle\langle\psi_i|$:
\begin{gather}
\mathcal{E}_S\big(|\psi_i\rangle\langle\psi_i|\big) = \cos^2\frac{\theta_i}{2}\; \mathcal{E}_S(\ket{1}\bra{1})+\sin^2\frac{\theta_i}{2}\; \mathcal{E}_S(\ket{0}\bra{0})
 \nonumber \\ +  e^{-i\phi_i}\cos\frac{\theta_i}{2}\sin\frac{\theta_i}{2}\; \mathcal{E}_S(\ket{1}\bra{0}) + \text{h.c.} 
\end{gather}
The GAD channel modifies the fundamental basis operators according to:
\begin{gather}
    \mathcal{E}_S(\ket{1}\bra{1}) = \big[1 - (1-p)\gamma\big]\ket{1}\bra{1} + (1-p)\gamma\ket{0}\bra{0},  \nonumber \\  
    \mathcal{E}_S(\ket{0}\bra{0}) = \gamma p\ket{1}\bra{1} + \big(1 - \gamma p\big)\ket{0}\bra{0},  \nonumber \\ 
    \mathcal{E}_S(\ket{1}\bra{0}) = \sqrt{1-\gamma}\ket{1}\bra{0},  
    \mathcal{E}_S(\ket{0}\bra{1}) = \sqrt{1-\gamma}\ket{0}\bra{1}. \label{eq:app_map_01}
\end{gather}
Substituting these transformations, we obtain:
\begin{gather}
\mathcal{E}_S\big(|\psi_i\rangle\langle\psi_i|\big) = \left[ \cos^2\frac{\theta_i}{2}\big[1 - (1-p)\gamma\big] + \sin^2\frac{\theta_i}{2}\gamma p \right] \ket{1}\bra{1}  \nonumber \\ + \left[ \sin^2\frac{\theta_i}{2}\big(1 - \gamma p\big) + \cos^2\frac{\theta_i}{2} (1-p)\gamma \right] \ket{0}\bra{0} \nonumber \\
\quad + \left[ e^{-i\phi_i}\cos\frac{\theta_i}{2}\sin\frac{\theta_i}{2}\sqrt{1-\gamma} \right] \ket{1}\bra{0} + \text{h.c.}
\end{gather}

We then compute the trace by sandwiching $\mathcal{E}_S\big(|\psi_i\rangle\langle\psi_i|\big)$ with the post-selected state $\ket{\psi_f}$. The inner products with the basis operators yield:
\begin{gather}
    \bra{\psi_f} \ket{1}\bra{1} \ket{\psi_f} = \cos^2\frac{\theta_f}{2}, \quad  \bra{\psi_f} \ket{0}\bra{0} \ket{\psi_f} = \sin^2\frac{\theta_f}{2}, \nonumber\\
    \bra{\psi_f} \ket{1}\bra{0} \ket{\psi_f} = e^{i\phi_f}\cos\frac{\theta_f}{2}\sin\frac{\theta_f}{2}, \nonumber\\  \bra{\psi_f} \ket{0}\bra{1} \ket{\psi_f} = e^{-i\phi_f}\cos\frac{\theta_f}{2}\sin\frac{\theta_f}{2}.
\end{gather}
Multiplying these projection weights by the corresponding terms in $\mathcal{E}_S\big(|\psi_i\rangle\langle\psi_i|\big)$, the denominator becomes:
\begin{gather}
\bra{\psi_f} \mathcal{E}_S\big(|\psi_i\rangle\langle\psi_i|\big) \ket{\psi_f} =  \nonumber\\ \cos^2\frac{\theta_f}{2} \left[ \cos^2\frac{\theta_i}{2}\big[1 - (1-p)\gamma\big] +   
\sin^2\frac{\theta_i}{2}\gamma p \right] \nonumber\\ + \sin^2\frac{\theta_f}{2} \left[ \sin^2\frac{\theta_i}{2}\big(1 - \gamma p\big) + \cos^2\frac{\theta_i}{2}(1-p)\gamma \right] \nonumber \\
\quad + 2\cos(\phi_i - \phi_f)\cos\frac{\theta_i}{2}\sin\frac{\theta_i}{2}\cos\frac{\theta_f}{2}\sin\frac{\theta_f}{2}\sqrt{1-\gamma}.  \label{eq:appendix_denominator}
\end{gather}

\subsection{Evaluation of the Numerator}
For the numerator, $\bra{\psi_f} \mathcal{E}_S\big(\hat{\sigma}_z |\psi_i\rangle\langle\psi_i|\big) \ket{\psi_f}$, the observable $\hat{\sigma}_z = \ket{1}\bra{1} - \ket{0}\bra{0}$ acts on the initial state before the interaction with the environment. Expanding this operation gives:
\begin{gather}
\label{eq:operator_O}
\hat{\sigma}_z |\psi_i\rangle\langle\psi_i| = \cos^2\frac{\theta_i}{2}\ket{1}\bra{1} - \sin^2\frac{\theta_i}{2}\ket{0}\bra{0} \nonumber \\
+ e^{-i\phi_i}\cos\frac{\theta_i}{2}\sin\frac{\theta_i}{2}\ket{1}\bra{0} - \text{h.c.}
\end{gather}
Notably, $\hat{\sigma}_z$ preserves the diagonal and off-diagonal structure of the density operator but inverts the signs of the ground-state population and the $\ket{0}\bra{1}$ coherence. Applying the channel transformations \eqref{eq:app_map_01} yields:
\begin{gather}
\mathcal{E}_S\big(\hat{\sigma}_z |\psi_i\rangle\langle\psi_i| \big) =   \nonumber \\ \left[ \cos^2\frac{\theta_i}{2}\big[1 - (1-p)\gamma\big] - \sin^2\frac{\theta_i}{2}\gamma p \right] \ket{1}\bra{1}  + \nonumber \\  \left[ \cos^2\frac{\theta_i}{2}(1-p)\gamma - \sin^2\frac{\theta_i}{2}\big(1 - \gamma p\big) \right] \ket{0}\bra{0} \nonumber \\
\quad + \left[ e^{-i\phi_i}\cos\frac{\theta_i}{2}\sin\frac{\theta_i}{2}\sqrt{1-\gamma} \right] \ket{1}\bra{0} - \text{h.c.}
\end{gather}

Sandwiching $\mathcal{E}_S\big(\hat{\sigma}_z |\psi_i\rangle\langle\psi_i| \big)$ with the post-selected state $\ket{\psi_f}$, the final expression for the numerator evaluates to:
\begin{gather}
\bra{\psi_f} \mathcal{E}_S\big(\hat{\sigma}_z |\psi_i\rangle\langle\psi_i|\big) \ket{\psi_f} =  \nonumber \\  \cos^2\frac{\theta_f}{2} \left[ \cos^2\frac{\theta_i}{2}\big[1 - (1-p)\gamma\big] - \sin^2\frac{\theta_i}{2}\gamma p \right]  +   \nonumber \\  \sin^2\frac{\theta_f}{2} \left[ \cos^2\frac{\theta_i}{2} (1-p)\gamma - \sin^2\frac{\theta_i}{2}\big(1 - \gamma p\big) \right] \nonumber \\
\quad - 2i\sin(\phi_i - \phi_f)\cos\frac{\theta_i}{2}\sin\frac{\theta_i}{2}\cos\frac{\theta_f}{2}\sin\frac{\theta_f}{2}\sqrt{1-\gamma}. \label{eq:appendix_numerator}
\end{gather}

The total operational weak value $\langle \hat{\sigma}_z \rangle_w^{\mathcal{E}}$ is precisely given by the ratio of Eq.~\eqref{eq:appendix_numerator} to Eq.~\eqref{eq:appendix_denominator}. To obtain a more physically transparent expression, we simplify the half-angle dependencies in both the numerator and the denominator by applying the double-angle identities: $\cos^2(\theta/2) = (1+\cos\theta)/2$, $\sin^2(\theta/2) = (1-\cos\theta)/2$, and $2\sin(\theta/2)\cos(\theta/2) = \sin\theta$. The weak value expression simplifies to:
\begin{widetext}
    \begin{gather}
    \langle \hat{\sigma}_z \rangle_w^{\mathcal{E}} = \frac{\cos\theta_i + \cos\theta_f \big[ 1 - \gamma + \gamma (2p-1) \cos\theta_i \big] - i\sin(\phi_i - \phi_f)\sin\theta_i\sin\theta_f\sqrt{1-\gamma}}{1 + \cos\theta_i\cos\theta_f(1-\gamma) + \gamma (2p-1)\cos\theta_f + \cos(\phi_i - \phi_f)\sin\theta_i\sin\theta_f\sqrt{1-\gamma}}.
\end{gather}
\end{widetext}

\section{Derivation of Success Probability}
\label{app:psucc_derivation}

The post-selection probability $P_{\text{succ}}$ is given by the trace of the unnormalized apparatus state $\hat{\rho}^{\text{ps}} = \sum_j \hat{\tilde{\rho}}_j$. Expressing this in terms of the unnormalized coefficients $\mu_j^{\pm} = \sqrt{P_{\text{succ}}} \, c_j^{\pm}$, the $j$-th component becomes:
\begin{gather}
\hat{\tilde{\rho}}_j = |\mu_{j}^{+}|^2 \hat{D}_{+} \hat{\rho}_{0} \hat{D}_{+}^\dagger 
    + |\mu_{j}^{-}|^2 \hat{D}_{-} \hat{\rho}_{0} \hat{D}_{-}^\dagger 
   \nonumber \\  + \mu_{j}^{+} (\mu_{j}^{-})^* e^{-ibs} \hat{D}_{+} \hat{\rho}_{0} \hat{D}_{-}^\dagger 
    + (\mu_{j}^{+})^* \mu_{j}^{-} e^{ibs} \hat{D}_{-} \hat{\rho}_{0} \hat{D}_{+}^\dagger . 
\end{gather}
The success probability is the sum of the traces of these components, $P_{\text{succ}} = \sum_j \mathrm{Tr}[\hat{\tilde{\rho}}_j]$. Evaluating the trace of the diagonal terms yields the simple population sum $|\mu_j^+|^2 + |\mu_j^-|^2$, since $\mathrm{Tr} [\hat{D}_{\nu} \hat{\rho}_{0} \hat{D}_{\nu}^\dagger ]=\mathrm{Tr} [ \hat{\rho}_{0}  ] =1$, with $\nu \in \{+,-\}$. 

For the off-diagonal interference terms, we evaluate the trace of the operator product $\hat{D}_{+} \hat{\rho}_{0} \hat{D}_{-}^\dagger$. Using the cyclic property of the trace and the Weyl composition rule [$\hat{D}_{-}^\dagger \hat{D}_{+} = e^{-ibs} \hat{D}(s)$], we obtain:
\begin{equation}\label{eq:chi_0}
    \mathrm{Tr}\left[ \hat{D}_{+} \hat{\rho}_{0} \hat{D}_{-}^\dagger \right] = e^{-ibs} \mathrm{Tr}[\hat{\rho}_0 \hat{D}(s)] = e^{-ibs} \chi_0(s),
\end{equation}
where $\chi_0(s) = \mathrm{Tr}[\hat{\rho}_0 \hat{D}(s)] = \exp(-\Gamma s^2)$ is the characteristic function of a thermal squeezed state~\cite{ferraro2005gaussianstatescontinuousvariable}. Here, the rate $\Gamma = (2\bar{n}+1) (\cosh 2r - \cos 2\chi \sinh 2r)$ is determined by the squeezing and thermal parameters of the probe. The total interference contribution is the product of this trace and the coefficient $\mu_{j}^{+} (\mu_{j}^{-})^* e^{-ibs}$. The phase factors from the coefficient and the operator trace combine constructively to yield:
\begin{equation}
    \mu_{j}^{+} (\mu_{j}^{-})^* e^{-ibs} \left[ e^{-ibs} \chi_0(s) \right] = \mu_{j}^{+} (\mu_{j}^{-})^* \left[ e^{-2ibs} \exp(-\Gamma s^2) \right].
\end{equation}
 Thus, the general formula for the probability becomes:
\begin{gather}
    P_{\text{succ}} = \nonumber \\ \sum_j \Bigg\{ |\mu_j^+|^2 + |\mu_j^-|^2 
  +2\mathrm{Re}\left[ \mu_j^+ (\mu_j^-)^*  e^{-2ibs} \exp(-\Gamma s^2) \right] \Bigg\}.
\end{gather}
Finally, by substituting the explicit GAD forms for $\mu_j^{\pm}$:
\begin{gather}
\mu_0^+ = \sqrt{(1-p)(1-\gamma)} \cos\frac{\theta_f}{2} \cos\frac{\theta_i}{2}, \nonumber \\
    \mu_0^- = \sqrt{(1-p)} e^{i(\phi_i-\phi_f)} \sin\frac{\theta_f}{2} \sin\frac{\theta_i}{2},  \nonumber \\
    \mu_1^+ = \sqrt{(1-p)\gamma} e^{-i\phi_f}\sin\frac{\theta_f}{2} \cos\frac{\theta_i}{2}, \;\;\;\;\;
    \mu_1^- = 0, \nonumber \\
    \mu_2^+ = \sqrt{p} \cos\frac{\theta_f}{2}\cos\frac{\theta_i}{2},   \nonumber \\
    \mu_2^- = \sqrt{p(1-\gamma)} e^{i(\phi_i-\phi_f)}\sin\frac{\theta_f}{2}\sin\frac{\theta_i}{2}, \nonumber \\
    \mu_3^+ = 0, \;\;\;\;\; 
    \mu_3^- = \sqrt{p\gamma} e^{i\phi_i} \cos\frac{\theta_f}{2} \sin\frac{\theta_i}{2}.
\end{gather}
and summing over all outcomes $j$, we recover the closed-form expression:
\begin{gather}
P_{\text{succ}} = \frac{1}{2} \Big[ 1 + \gamma (2p-1)\cos\theta_f + (1-\gamma)\cos\theta_i\cos\theta_f \Big] \nonumber \\
    + \frac{e^{-\Gamma s^2}}{2} \sqrt{1-\gamma} \sin\theta_i \sin\theta_f \cos(2bs + \phi_i - \phi_f).
\end{gather}

\section{Gaussian Identities}

The exact analytical evaluation of the quantum overlap between the conditioned and initial apparatus states relies on computing the trace of shifted phase-space distributions. Because the measurement apparatus is initialized in a Gaussian state, these trace operations can be exactly resolved using specific Gaussian identities. By considering the centered squeezed thermal reference state $\hat{\rho}_0$, these identities can be derived following a method based on the integration of characteristic functions, similar to that in Ref.~\cite{MarianPRA1993}.

The first identity governs the overlap between the reference state $\hat{\rho}_0$ and its displaced copy, $\hat{D}(s)\hat{\rho}_0 \hat{D}^\dagger(s)$:
\begin{equation}\label{eq:overlap_identity}
\mathrm{Tr}\!\left[
\hat{\rho}_0 \hat{D}(s)\hat{\rho}_0 \hat{D}^\dagger(s)
\right]
=
\mathcal{P}_0\,\exp\!\left(-2\widetilde{\Gamma} s^2\right),
\end{equation}
where $\mathcal{P}_0 \equiv \mathrm{Tr}(\hat{\rho}_0^2)=(2\bar n+1)^{-1}$ represents the purity of the reference state~\cite{MarianPRA1993,serafini2017quantum}. The decay envelope is determined by $\widetilde{\Gamma}=\frac{1}{2\bar n+1} \left( \cosh 2r - \cos 2\chi\,\sinh 2r \right)$, which characterizes the inverse covariance of $\hat{\rho}_0$. This term describes the classical spatial overlap, which decays slowly for states with broad spatial support (such as anti-squeezed states).

The second identity involves the trace of squared coherence terms, which corresponds to displacements in the same direction. For Gaussian states displaced along a principal axis, this term shares the same decay profile as the population overlap, governed by the inverse covariance:
\begin{equation}\label{eq:identity_phase}
\mathrm{Tr}\!\left[
\hat{\rho}_0 \hat{D}(s) \hat{\rho}_0 \hat{D}(s)
\right]
=
\mathcal{P}_0 \exp\!\left(-2\widetilde{\Gamma} s^2\right).
\end{equation}
Physically, both identities originate from the spatial overlap of the wavepacket with its shifted version. Since the probe state is prepared in a broad configuration to facilitate weak measurements, both the population and interference terms persist over a large range of $s$, decaying at the slow rate $\widetilde{\Gamma}$.

\section{Quantum Overlap}
\label{app:overlap_explicit}

The quantum overlap $\mathcal{O} = \mathrm{Tr}(\hat{\rho}_{M} \hat{\rho}_M^{\mathrm{ps}})$ quantifies the fidelity between the initial apparatus state and its final conditioned state resulting from the system's post-selection. To evaluate this analytically, we exploit the unitary invariance of the trace to shift the coordinate system to the origin of phase space. The initial apparatus state is defined as $\hat{\rho}_{M} = \hat{D}(\alpha) \hat{\rho}_{0} \hat{D}^\dagger(\alpha)$, where $\hat{\rho}_{0}$ is the centered squeezed thermal reference state. The conditioned state $\hat{\rho}_M^{\mathrm{ps}}$ is a sum of components $\hat{\rho}_{\lambda\nu}$ weighted by coefficients $\mathcal{W}_{\lambda\nu}$. The overlap can therefore be decomposed into four traces:
\begin{equation}
    \mathcal{O} = \mathcal{W}_{++} T_{++} + \mathcal{W}_{--} T_{--} + e^{-ibs}\mathcal{W}_{+-} T_{+-} + e^{ibs}\mathcal{W}_{-+} T_{-+},
\end{equation}
where $T_{\lambda\nu} = \mathrm{Tr}[\hat{\rho}_M \hat{\rho}_{\lambda\nu}]$. Using the Weyl relations $\hat{D}^\dagger(\alpha)\hat{D}(\beta) = e^{i\mathrm{Im}(\alpha\beta^*)}\hat{D}(\beta-\alpha)$, we map the problem to overlaps involving the centered reference state $\hat{\rho}_0$. For the diagonal terms, $T_{++}$ and $T_{--}$, the Weyl phases cancel out. The trace reduces to the overlap of the reference state with a copy displaced by $\pm s/2$, given by:
\begin{equation}
    T_{++} = T_{--} = \mathrm{Tr}[\hat{\rho}_0 \hat{D}(s/2) \hat{\rho}_0 \hat{D}^\dagger(s/2)] = \mathcal{P}_0 \exp\left( -\frac{ \widetilde{\Gamma}s^2}{2} \right),
\end{equation}
where we used identity~(\ref{eq:overlap_identity}).

For the off-diagonal terms, $T_{+-}$ and $T_{-+}$, the Weyl phases accumulate rather than cancel. The trace reduces to the self-interference of the reference state:
\begin{align}
    T_{+-} &= e^{-ibs} \mathrm{Tr}[\hat{\rho}_0 \hat{D}(s/2) \hat{\rho}_0 \hat{D}(s/2)] = e^{-ibs} \mathcal{P}_0 \exp\left( -\frac{\widetilde{\Gamma} s^2}{2} \right), \\
    T_{-+} &= e^{ibs} \mathcal{P}_0 \exp\left( -\frac{\widetilde{\Gamma}s^2}{2} \right),
\end{align}
where we used the identity~(\ref{eq:identity_phase}). Note that both the diagonal and off-diagonal traces are governed by the same inverse covariance rate $\widetilde{\Gamma}$, as they both originate from the spatial overlap of the broad pointer wavepacket.

Combining these traces with the state coefficients, the phase factors in the interference term accumulate ($e^{-ibs}$ from the coefficient and $e^{-ibs}$ from the trace), yielding the total phase $e^{-2ibs}$. The exact analytic overlap is:
\begin{equation}
\label{eq:overlap_final}
\mathcal{O} = \mathcal{P}_0 \, e^{-\frac{\widetilde{\widetilde{\Gamma}} s^2}{2}} \left[
\Omega_1
+
\Omega_2  
\right],
\end{equation}
where the weights are defined as:
\begin{subequations}
\begin{align}
\Omega_1 &= \mathcal{W}_{++} + \mathcal{W}_{--} = \sum_j \left(|c_j^+|^2 + |c_j^-|^2\right), \\
\Omega_2 &= 2\mathrm{Re}\left( \mathcal{W}_{+-}  e^{-2ibs} \right) = 2\sum_j \mathrm{Re}\left[  c_j^+ (c_j^-)^*\;e^{-2ibs} \right] .
\end{align}
\end{subequations}
Here, $\Omega_1$ represents the classical population contribution, while $\Omega_2$ captures the quantum interference. Both contributions share the same Gaussian decay envelope governed by $\widetilde{\Gamma}$. However, the interference term is distinguished by its phase sensitivity, oscillating with the initial momentum $b$ via the factor $e^{-2ibs}$. Substituting the explicit GAD coefficients $c_j^{\pm}$ (see Appendix~\ref{app:GAD-coef}) we obtain:
\begin{equation}
\label{eq:overlap_simplified_ratio}
\mathcal{O}(s) = \frac{ e^{-\frac{\widetilde{\Gamma} s^2}{2}} \left[ \eta + \Psi \right]}
{(2\bar{n}+1)(\eta + e^{-\Gamma s^2} \Psi)},
\end{equation}
where
\begin{gather}
     \eta = 1 + \gamma (2p-1)\cos\theta_f + (1-\gamma)\cos\theta_i\cos\theta_f, \nonumber \\  \Psi = \sqrt{1-\gamma}\sin\theta_i\sin\theta_f\cos(2bs + \phi_i - \phi_f).
\end{gather}

\section{Appendix: Calculation of the Mean Position Shift, Variance, and Signal-to-Noise Ratio}\label{app:shift_X}

In this section, we derive the net spatial shift of the pointer, $\Delta X$, induced by the weak measurement interaction. Because the interaction Hamiltonian is translationally invariant with respect to the initial pointer position, the resulting shift is independent of the initial spatial coordinate. Therefore, without loss of generality, we can set the initial mean position to the origin ($x_0 = 0$). This corresponds to an initial phase-space displacement that is purely imaginary, $\alpha = ib$. Under this simplification, the net position shift is simply the final expectation value, $\Delta X = \text{Tr}(\hat{X} \rho_M^{PS})$. Using the linearity of the trace, we evaluate the expectation value separately for the diagonal and off-diagonal components of the conditioned state.

\subsection{Diagonal Terms}

We first evaluate the trace for the diagonal component $\rho_{++}$:
\begin{equation}
    \text{Tr}(\hat{X} \rho_{++}) = \text{Tr}(\hat{X} \hat{D}_+ \rho_0 \hat{D}_+^\dagger).
\end{equation}
Using the cyclic property of the trace, this becomes $\text{Tr}(\hat{D}_+^\dagger \hat{X} \hat{D}_+ \rho_0)$. The Heisenberg evolution of the position operator under a general displacement $\hat{D}(\gamma)$ is given by $\hat{D}^\dagger(\gamma) \hat{X} \hat{D}(\gamma) = \hat{X} + 2\sigma\text{Re}(\gamma)$, where $\sigma$ is the spatial width of the pointer. Inserting the post-interaction displacement $\gamma = ib + s/2$, the operator transforms as:
\begin{equation}
    \hat{D}_+^\dagger \hat{X} \hat{D}_+ = \hat{X} + 2\sigma\text{Re}\left(ib + \frac{s}{2}\right) = \hat{X} + \sigma s.
\end{equation}
Because the initial thermal squeezed state $\rho_0$ is centered at the origin, its first moment identically vanishes, $\text{Tr}(\hat{X}\rho_0) = 0$. Therefore, the trace immediately yields the classical spatial shift:
\begin{equation}
    \text{Tr}(\hat{X} \rho_{++}) = \text{Tr}\left[ \left(\hat{X} + \sigma s\right) \rho_0 \right] = \sigma s.
\end{equation}
By analogy, for the negative displacement component ($\gamma = ib - s/2$), we obtain:
\begin{equation}
    \text{Tr}(\hat{X} \rho_{--}) = \text{Tr}\left[ \left(\hat{X} - \sigma s\right) \rho_0 \right] = -\sigma s.
\end{equation}

\subsection{Off-Diagonal Terms}

Next, we evaluate the contribution of the cross-term $\rho_{+-}$:
\begin{equation}
    \text{Tr}(\hat{X} \rho_{+-}) = \text{Tr}(\hat{X} \hat{D}_+ \rho_0 \hat{D}_-^\dagger).
\end{equation}
Since $x_0 = 0$, the initial displacements are purely imaginary: $\hat{D}_\pm = \hat{D}(ib \pm s/2)$. Using the Weyl displacement relation, we factor out the initial momentum kick as $\hat{D}_\pm = \hat{D}(ib) \hat{D}(\pm s/2) e^{\mp i b s/2}$. Substituting this factorization into the trace yields:
\begin{equation}
    \text{Tr}(\hat{X} \rho_{+-}) = e^{-isb} \text{Tr}\left[ \hat{X} \hat{D}(ib) \hat{D}(s/2) \rho_0 \hat{D}(s/2) \hat{D}^\dagger(ib) \right].
\end{equation}
Using the cyclic property of the trace, we move $\hat{D}^\dagger(ib)$ to the far left. Because $ib$ is purely imaginary, its real part is zero, leaving the position operator invariant under this displacement: $\hat{D}^\dagger(ib) \hat{X} \hat{D}(ib) = \hat{X} + 2\sigma\text{Re}(ib) = \hat{X}$. Thus, the trace isolates the spatial shift:
\begin{equation}
    \text{Tr}(\hat{X} \rho_{+-}) = e^{-isb} \text{Tr}\left[ \hat{X} \hat{D}(s/2) \rho_0 \hat{D}(s/2) \right].
\end{equation}

To prove that this trace is null, we apply the cyclic property to rewrite it as $\text{Tr}\left[ \hat{D}(s/2) \hat{X} \hat{D}(s/2) \rho_0 \right]$. Using the operator identity $\hat{D}(s/2) \hat{X} \hat{D}(s/2) = \hat{D}(s) (\hat{X} + \sigma s)$, where we have inserted the identity $\hat{D}(s/2) \hat{D}^{\dagger}(s/2) = \hat{\mathbbm{1}}$ before the $\hat{X}$ operator, we expand it into two traces:
\begin{equation} \label{eq:J_expansion}
     \text{Tr}\left[ \hat{D}(s) \hat{X} \rho_0 \right] + \sigma s \text{Tr}\left[ \hat{D}(s) \rho_0 \right].
\end{equation}
We evaluate the first trace in the continuous position basis $|x\rangle$. The spatial translation operator acts to the left as $\langle x | \hat{D}(s) = \langle x - 2\sigma s |$. Applying the position operator yields:
\begin{equation}
    \text{Tr}\left[ \hat{D}(s) \hat{X} \rho_0 \right] = \int_{-\infty}^\infty dx (x - 2\sigma s) \langle x - 2\sigma s | \rho_0 | x \rangle.
\end{equation}
By shifting the integration variable to symmetrically center the shift, $y = x - \sigma s$, the integral becomes:
\begin{equation}
    \int_{-\infty}^\infty dy (y - \sigma s) \langle y - \sigma s | \rho_0 | y + \sigma s \rangle.
\end{equation}
Because the initial state $\rho_0$ is centered symmetrically at the origin, its matrix element $\langle y - \sigma s | \rho_0 | y + \sigma s \rangle$ is strictly an even function of $y$. Consequently, the integral over the odd variable $y$ vanishes exactly. We are left solely with the constant term evaluated over the full trace:
\begin{equation}\label{eq:usefull_identity}
    -\sigma s \int_{-\infty}^\infty dy \langle y - \sigma s | \rho_0 | y + \sigma s \rangle = -\sigma s \text{Tr}[ \hat{D}(s) \rho_0 ].
\end{equation}
Substituting this directly back into Eq.~\eqref{eq:J_expansion}, we immediately see that the terms perfectly cancel:
\begin{equation}
     -\sigma s \text{Tr}[ \hat{D}(s) \rho_0 ] + \sigma s \text{Tr}[ \hat{D}(s) \rho_0 ] = 0.
\end{equation}
Therefore, $\text{Tr}(\hat{X} \rho_{+-}) = 0$. By identical symmetry, the conjugate cross-term $\rho_{-+}$ also reaches zero. This strictly proves that the quantum interference terms do not contribute to the position shift.

\subsection{Total Mean Position Shift}

Because the off-diagonal terms vanish identically in the position basis, the net spatial shift of the pointer is governed entirely by the diagonal elements. By summing these contributions, weighted by their respective coefficients, we arrive at the final analytical result:
\begin{equation}
    \Delta X = \sigma s (\mathcal{W}_{++} - \mathcal{W}_{--}).
\end{equation}
The quantity $\sigma s$ represents the fundamental spatial shift of the pointer when perfectly correlated with a definite system eigenstate, yielding a shift of $+\sigma s$ when the system is in the $|1\rangle$ eigenstate, and $-\sigma s$ when the system is in the $|0\rangle$ eigenstate. Consequently, the mean position shift can be factored as $\Delta X = (\sigma s) \mathcal{A}_X(s)$, where we define the dimensionless amplification factor:
\begin{equation}
    \mathcal{A}_X(s) \equiv \mathcal{W}_{++} - \mathcal{W}_{--}.
\end{equation}
Substituting the explicit GAD coefficients $c_j^{\pm}$ (see Appendix~\ref{app:GAD-coef}), we obtain:
\begin{equation}
\label{eq:amplification_z_appx}
    \mathcal{A}_X(s) = \frac{1}{2 P_{\text{succ}}} \Big\{ \cos\theta_i + \cos\theta_f \big[ 1 - \gamma + \gamma (2p-1) \cos\theta_i \big] \Big\}.
\end{equation}

Now, we examine the asymptotic limit as $s \to \infty$ of Eq.~\eqref{eq:amplification_z_appx} to verify the transition from an anomalous weak value to the conditional expectation value of the observable in a strong, post-selected projective measurement. As established in Eq.~\eqref{eq:Psucc_simplified}, the post-selection success probability contains a quantum interference cross-term modulated by the Gaussian envelope $e^{-\Gamma s^2}$ that goes to zero under this limit. Consequently, the success probability reduces to its fully decohered, classical transition counterpart:
\begin{equation}
\label{eq:Psucc_infinity}
    \lim_{s \to \infty} P_{\text{succ}}(s) = \frac{1}{2} \Big[ 1 + \gamma(2p-1)\cos\theta_f + (1-\gamma)\cos\theta_i\cos\theta_f \Big].
\end{equation}
Substituting Eq.~\eqref{eq:Psucc_infinity} back into Eq.~\eqref{eq:amplification_z_appx} yields the strong-measurement amplification asymptote:
\begin{equation}
\label{eq:AX_infinity}
    \lim_{s \to \infty} \mathcal{A}_X(s) = \frac{\cos\theta_i + \cos\theta_f \big[ 1 - \gamma + \gamma (2p-1) \cos\theta_i \big]}{1 + \gamma(2p-1)\cos\theta_f + (1-\gamma)\cos\theta_i\cos\theta_f}.
\end{equation}
We now demonstrate that Eq.~\eqref{eq:AX_infinity} is precisely the classical conditional expectation value $\langle \hat{\sigma}_z \rangle_{\text{cond}}$ of the system observable. In an ideal strong measurement, the apparatus completely destroys the coherence of the initial state $|\psi_i\rangle$, projecting it into the definite system eigenstates $|1\rangle$ and $|0\rangle$ with prior probabilities $p(1|i) = \frac{1}{2}(1 + \cos\theta_i)$ and $p(0|i) = \frac{1}{2}(1 - \cos\theta_i)$, respectively. These eigenstates components subsequently traverse the GAD channel $\mathcal{E}$, which contracts an arbitrary input  $\langle \hat{\sigma}_z \rangle_{\text{in}}$ according to the affine map $\langle \hat{\sigma}_z \rangle_{\text{out}} \to (1-\gamma)\langle \hat{\sigma}_z \rangle_{\text{in}} + \gamma(2p-1)$. Evaluating this transformation for the  system eigenstates- namely, substituting $\langle 1|\hat{\sigma}_z|1\rangle = +1$ and $\langle 0|\hat{\sigma}_z|0\rangle = -1$-explicitly yields:
\begin{equation}
    \langle \hat{\sigma}_z \rangle_{1}^{\mathcal{E}} = 1 - \gamma + \gamma(2p-1), \quad \quad \langle \hat{\sigma}_z \rangle_{0}^{\mathcal{E}} = -1 + \gamma + \gamma(2p-1).
\end{equation}
The subsequent probability of successfully post-selecting the state $|\psi_f\rangle$, conditioned on the intermediate projection into eigenstate $k \in \{1, 0\}$, is given by $p(f|k) = \frac{1}{2} \big(1 + \langle \hat{\sigma}_z \rangle_{k}^{\mathcal{E}} \cos\theta_f \big)$. Applying Bayes' theorem, the conditional expectation value of the pointer reading is:
\begin{equation}
\label{eq:bayes_cond_exp}
    \langle \hat{\sigma}_z \rangle_{\text{cond}} = \frac{(+1) p(f|1)p(1|i) + (-1) p(f|0)p(0|i)}{p(f|1)p(1|i) + p(f|0)p(0|i)}.
\end{equation}
Expanding the numerator of Eq.~\eqref{eq:bayes_cond_exp} yields:
\begin{gather}
    \text{Num} = \frac{1}{4} \Big[ (1+\cos\theta_i)\big(1 + \langle \hat{\sigma}_z \rangle_{1}^{\mathcal{E}}\cos\theta_f\big)  \nonumber \\ - (1-\cos\theta_i)\big(1 + \langle \hat{\sigma}_z \rangle_{0}^{\mathcal{E}}\cos\theta_f\big) \Big] \nonumber \\
    = \frac{1}{2} \Big\{ \cos\theta_i + \cos\theta_f \big[ 1 - \gamma + \gamma(2p-1)\cos\theta_i \big] \Big\},
\end{gather}
while expanding the denominator yields exactly the asymptotic success probability of Eq.~\eqref{eq:Psucc_infinity}. Dividing the numerator by the denominator establishes the identity:
\begin{equation}
    \lim_{s \to \infty} \mathcal{A}_X(s) = \langle \hat{\sigma}_z \rangle_{\text{cond}}.
\end{equation}
This confirms that as the measurement strength scales to infinity, the continuous pointer shift governed by our open-system model converges to the classical conditional expectation value dictated by the non-unitary dynamics of the GAD channel.


\subsection{Variance and Signal-to-Noise Ratio}\label{app:SNR}

In order to quantify the metrological advantage of the framework, we analyze the Signal-to-Noise Ratio (SNR) of the spatial shift. Following the established benchmark for post-selected weak measurements, the SNR for an estimator based on the average measurement result over $N$ trials is proportional to the mean shift and inversely bounded by the root-variance of the pointer~\cite{PangPRL2015}~:
\begin{equation}
\label{eq:SNR_apendix}
    \text{SNR} = \frac{\sqrt{N P_{\text{succ}}} |\Delta X|}{\sqrt{\text{Var}(\hat{X})}} = \frac{\sqrt{N P_{\text{succ}}} \sigma s |\mathcal{A}_X(s)|}{\sqrt{\text{Var}(\hat{X})}},
\end{equation}
where $\text{Var}(\hat{X}) = \text{Tr}(\hat{X}^2 \hat{\rho}_M^{\text{ps}}) - (\Delta X)^2$ is the variance of the pointer evaluated on the final, conditioned state. We begin by evaluating the expectation value of the second moment, $\langle \hat{X}^2 \rangle$. Let $\sigma_0^2 = \text{Tr}(\hat{X}^2 \hat{\rho}_0)= \sigma^2 (2\bar{n} + 1) \left( \cosh 2r - \sinh 2r \cos 2\chi \right)$ denote the initial spatial variance of the thermal squeezed pointer state. 
For the diagonal component $\hat{\rho}_{++} = \hat{D}_+ \hat{\rho}_0 \hat{D}_+^\dagger$, the Heisenberg evolution of $\hat{X}^2$ yields:
\begin{gather}
    \text{Tr}(\hat{X}^2 \hat{\rho}_{++}) = \text{Tr}(\hat{D}_+^\dagger \hat{X}^2 \hat{D}_+ \hat{\rho}_0) = \text{Tr}\left[ (\hat{X} + \sigma s)^2 \hat{\rho}_0 \right] =  \nonumber \\ \text{Tr}(\hat{X}^2 \hat{\rho}_0) + 2\sigma s \text{Tr}(\hat{X} \hat{\rho}_0) + \sigma^2 s^2 \text{Tr}(\hat{\rho}_0) = \sigma_0^2 + \sigma^2 s^2,
\end{gather}
where we used the fact that the initial state is centered at the origin, $\text{Tr}(\hat{X} \hat{\rho}_0) = 0$, and the purely imaginary initial momentum displacement $ib$ has a zero real part, yielding a classical spatial shift of  $\sigma s$. By symmetry, the negative component yields $\text{Tr}(\hat{X}^2 \hat{\rho}_{--}) = \sigma_0^2 + \sigma^2 s^2$. Next, we evaluate the off-diagonal cross-term $\hat{\rho}_{+-}$:
\begin{equation}
    \text{Tr}(\hat{X}^2 \hat{\rho}_{+-}) = \text{Tr}(\hat{X}^2 \hat{D}_+ \hat{\rho}_0 \hat{D}_-^\dagger).
\end{equation}
Using the Weyl displacement relation, we factor out the purely imaginary initial momentum as $\hat{D}_\pm = \hat{D}(ib)\hat{D}(\pm s/2) e^{\mp i b s / 2}$. Substituting this factorization into the trace explicitly extracts the momentum-dependent phase factor:
\begin{equation}
    \text{Tr}(\hat{X}^2 \hat{\rho}_{+-}) = e^{-isb}\; \text{Tr}\left[ \hat{X}^2 \hat{D}(ib) \hat{D}(s/2) \hat{\rho}_0 \hat{D}(s/2) \hat{D}^\dagger(ib) \right].
\end{equation}
Using the cyclic property of the trace, we move $\hat{D}^\dagger(ib)$ to the far left. Because $ib$ is purely imaginary, its real part is zero, leaving the position operator invariant under this specific displacement: $\hat{D}^\dagger(ib) \hat{X}^2 \hat{D}(ib) = \hat{X}^2$. Applying the remaining operator identity $\hat{D}(s/2) \hat{X}^2 \hat{D}(s/2) = \hat{D}(s) (\hat{X} + \sigma s)^2$, we expand the trace into three terms:
\begin{gather} \label{eq:X2_cross_expansion}    e^{-isb} \Big\{ \text{Tr}\left[ \hat{D}(s) \hat{X}^2 \hat{\rho}_0 \right] + 2\sigma s \text{Tr}\left[ \hat{D}(s) \hat{X} \hat{\rho}_0 \right] \nonumber \\ + \sigma^2 s^2 \text{Tr}\left[ \hat{D}(s) \hat{\rho}_0 \right] \Big\}.
\end{gather}
From our earlier derivation~\eqref{eq:usefull_identity}, we established that $\text{Tr}[\hat{D}(s)\hat{X}\hat{\rho}_0] = -\sigma s \text{Tr}[\hat{D}(s)\hat{\rho}_0]$. Therefore, the middle term evaluates to $-2\sigma^2 s^2 \text{Tr}[\hat{D}(s)\hat{\rho}_0]$. To evaluate the first term, we apply the position basis translation:
\begin{equation}
    \text{Tr}\left[ \hat{D}(s) \hat{X}^2 \hat{\rho}_0 \right] = \int_{-\infty}^\infty dy (y - \sigma s)^2 \langle y - \sigma s | \hat{\rho}_0 | y + \sigma s \rangle.
\end{equation}
Expanding $(y - \sigma s)^2 = y^2 - 2\sigma s y + \sigma^2 s^2$, the odd linear term $y$ vanishes upon integration against the even symmetric overlap function. This leaves:
\begin{equation}
    \int_{-\infty}^\infty dy y^2 \langle y - \sigma s | \hat{\rho}_0 | y + \sigma s \rangle + \sigma^2 s^2 \text{Tr}[ \hat{D}(s) \hat{\rho}_0 ].
\end{equation}
A fundamental property of zero-mean Gaussian states is that the spatial variance evaluated over this symmetrically shifted overlap integral is invariant under the shift $s$. Consequently, $\int dy y^2 \langle y - \sigma s | \hat{\rho}_0 | y + \sigma s \rangle = \sigma_0^2 \text{Tr}[ \hat{D}(s) \hat{\rho}_0 ]$. Substituting this back into Eq.~\eqref{eq:X2_cross_expansion}, the $\sigma^2 s^2$ terms perfectly cancel, yielding:
\begin{equation}
    \text{Tr}(\hat{X}^2 \hat{\rho}_{+-}) = \sigma_0^2 e^{-isb} \text{Tr}[ \hat{D}(s) \hat{\rho}_0 ].
\end{equation}
At this stage, we identify $\mathrm{Tr}[\hat{\rho}_0 \hat{D}(s)] = \exp(-\Gamma s^2)$ as the characteristic function of the thermal squeezed state~\cite{ferraro2005gaussianstatescontinuousvariable} (see Appedix~\ref{app:psucc_derivation}), where $\Gamma = (2\bar{n}+1) (\cosh 2r - \cos 2\chi \sinh 2r)$. Substituting this Gaussian envelope yields:
\begin{equation}
    \text{Tr}(\hat{X}^2 \hat{\rho}_{+-}) = \sigma_0^2 e^{-isb} e^{-\Gamma s^2}.
\end{equation}
Similarly, the conjugate cross-term evaluates to $\text{Tr}(\hat{X}^2 \hat{\rho}_{-+}) = \sigma_0^2 e^{isb} e^{-\Gamma s^2}$. To obtain the total expectation value $\langle \hat{X}^2 \rangle$, we weight the individual operator traces by their coefficients and corresponding phase factors. Summing these terms yields:
\begin{gather}
    \langle \hat{X}^2 \rangle = \mathcal{W}_{++}(\sigma_0^2 + \sigma^2 s^2) + \mathcal{W}_{--}(\sigma_0^2 + \sigma^2 s^2)  \nonumber \\ + e^{-ibs} \mathcal{W}_{+-} (\sigma_0^2 e^{-isb} e^{-\Gamma s^2}) + e^{ibs} \mathcal{W}_{-+} (\sigma_0^2 e^{isb} e^{-\Gamma s^2}) \nonumber \\
    = \sigma_0^2 \Big[ \mathcal{W}_{++} + \mathcal{W}_{--} + e^{-\Gamma s^2} \left( e^{-2ibs} \mathcal{W}_{+-} + e^{2ibs} \mathcal{W}_{-+} \right) \Big]\nonumber \\  + \sigma^2 s^2 (\mathcal{W}_{++} + \mathcal{W}_{--}).
\end{gather}
Because the coefficients and phase-modulated interference terms in the bracketed expression constitute the normalization of the conditioned state, the term evaluates to unity. The second moment simplifies to:
\begin{equation}
    \langle \hat{X}^2 \rangle = \sigma_0^2 + \sigma^2 s^2 (\mathcal{W}_{++} + \mathcal{W}_{--}).
\end{equation}
The variance is $\text{Var}(\hat{X}) = \langle \hat{X}^2 \rangle - (\Delta X)^2$. Substituting $\Delta X = \sigma s (\mathcal{W}_{++} - \mathcal{W}_{--})$, we find:
\begin{equation}
    \text{Var}(\hat{X}) = \sigma_0^2 + \sigma^2 s^2 \left[ (\mathcal{W}_{++} + \mathcal{W}_{--}) - (\mathcal{W}_{++} - \mathcal{W}_{--})^2 \right].
\end{equation}
Substituting the explicit GAD coefficients $c_j^{\pm}$ (see Appendix~\ref{app:GAD-coef}), we obtain: 
\begin{equation}
\label{eq:exact_variance_appendix}
    \text{Var}(\hat{X})_{PS} = \sigma^2 \Gamma + \sigma^2 s^2 \left[ \frac{A \left( A + B e^{-\Gamma s^2} \right) - C^2}{\left( A + Be^{-\Gamma s^2} \right)^2} \right],
\end{equation}
where we defined the following auxiliary variables:
\begin{gather}
    A = 1 + \cos\theta_f \big[ (1-\gamma)\cos\theta_i + \gamma(2p-1) \big], \nonumber \\
    B = \sqrt{1-\gamma} \sin\theta_i \sin\theta_f \cos(2bs - \phi_f + \phi_i),  \nonumber\\
    C = \cos\theta_i + \cos\theta_f \big[ 1 - \gamma + \gamma(2p-1)\cos\theta_i \big].
\end{gather}


\section{Calculation of the Mean Momentum Shift}\label{app:shift_P}

Now, we evaluate the net momentum shift of the pointer, defined as the difference between the final expectation value and the initial momentum: $\Delta P = \langle P \rangle_f - P_0 = \text{Tr}(\hat{P} \rho_M^{\text{ps}}) - P_0$. We keep the initial condition of a pointer centered at the origin ($x_0 = 0$), corresponding to a purely imaginary phase-space displacement $\alpha = ib$. Since the momentum operator is $\hat{P} = \frac{i}{2\sigma}(\hat{a}^{\dagger} - \hat{a})$, its Heisenberg evolution under a general displacement $\hat{D}(\gamma)$ is given by $\hat{D}^\dagger(\gamma) \hat{P} \hat{D}(\gamma) = \hat{P} + \text{Im}(\gamma)/\sigma$. For our initial state, this yields an initial mean momentum $P_0 = b/\sigma$. Considering this, we evaluate the diagonal and off-diagonal components of the final momentum $\langle P \rangle_f$ separately.

\subsection{Diagonal Terms}

We first evaluate the trace for the diagonal component $\rho_{++}$:
\begin{equation}
    \text{Tr}(\hat{P} \rho_{++}) = \text{Tr}(\hat{P} \hat{D}_+ \rho_0 \hat{D}_+^\dagger) = \text{Tr}(\hat{D}_+^\dagger \hat{P} \hat{D}_+ \rho_0).
\end{equation}
Simplifying the Heisenberg evolution $\hat{D}_+^\dagger \hat{P} \hat{D}_+$, and substituting $\gamma = ib + s/2$, we obtain
\begin{equation}
    \hat{D}_+^\dagger \hat{P} \hat{D}_+ = \hat{P} + \frac{\text{Im}\left(ib + s/2\right)}{\sigma} = \hat{P} + \frac{b}{\sigma} = \hat{P} + P_0.
\end{equation}
The trace simplifies to:
\begin{equation}
    \text{Tr}(\hat{P} \rho_{++}) = \text{Tr}\left[ \left(\hat{P} + P_0\right) \rho_0 \right] = P_0,
\end{equation}
where we have used $\text{Tr}(\hat{P}\rho_0) = 0$, since the zero-mean thermal squeezed state $\rho_0$ is centered at the origin of phase space. By exact analogy, the negative displacement component ($\gamma = ib - s/2$) yields the same imaginary part, resulting in $\text{Tr}(\hat{P} \rho_{--}) = P_0$. Thus, the diagonal terms induce no momentum shift; they simply return the initial momentum $P_0$.

\subsection{Off-Diagonal Terms}

Next, we evaluate the contribution of the cross-term $\rho_{+-}$:
\begin{equation}
    \text{Tr}(\hat{P} \rho_{+-}) = \text{Tr}(\hat{P} \hat{D}_+ \rho_0 \hat{D}_-^\dagger).
\end{equation}
We factor the total displacements using the Weyl relation, $\hat{D}_\pm = \hat{D}(ib) \hat{D}(\pm s/2) e^{\mp i b s/2}$. Substituting these and applying the cyclic property to move $\hat{D}^\dagger(ib)$ to the far left yields:
\begin{gather}
   \text{Tr}(\hat{P} \rho_{+-}) = \nonumber \\ \text{Tr}\left( \hat{P} \Big[ \hat{D}(ib) \hat{D}(s/2) e^{-ibs/2} \Big] \rho_0 \Big[ e^{-ibs/2} \hat{D}(s/2) \hat{D}^\dagger(ib) \Big] \right)  =  \nonumber \\ e^{-isb} \text{Tr}\left[ \hat{D}^\dagger(ib) \hat{P} \hat{D}(ib) \hat{D}(s/2) \rho_0 \hat{D}(s/2) \right].
\end{gather}
Applying the displacement rule, $\hat{D}^\dagger(ib) \hat{P} \hat{D}(ib) = \hat{P} + P_0$. We expand this into two distinct traces:
\begin{equation} \label{eq:P_offdiag_split}
    e^{-isb} P_0 \text{Tr}\left[ \hat{D}(s/2) \rho_0 \hat{D}(s/2) \right] + e^{-isb} \text{Tr}\left[ \hat{P} \hat{D}(s/2) \rho_0 \hat{D}(s/2) \right].
\end{equation}
Because the displacement parameter $s/2$ is purely real, the operator $\hat{D}(s/2)$ can be written as $\hat{D}(s/2) = \exp\left[ \frac{s}{2}(\hat{a}^\dagger - \hat{a}) \right]=  \exp(-is\sigma\hat{P})$, and therefore it commutes with $\hat{P}$. This commutation allows us to move $\hat{P}$ to the left and directly combine the half-displacements into $\hat{D}(s)$. To evaluate the remaining trace $\text{Tr}[ \hat{P} \hat{D}(s) \rho_0 ]$, we leverage the known characteristic function of the thermal squeezed state from Eq.~\eqref{eq:chi_0}, $\chi_0(s) = \text{Tr}[ \hat{D}(s) \rho_0 ] = e^{-\Gamma s^2}$. Taking the partial derivative of the displacement operator $\hat{D}(s) = \exp[s(\hat{a}^\dagger - \hat{a})]$ with respect to $s$ gives $\frac{\partial}{\partial s} \hat{D}(s) = (\hat{a}^\dagger - \hat{a}) \hat{D}(s)$. Using our definition of the momentum operator, we rewrite this as $(\hat{a}^\dagger - \hat{a}) = -2i\sigma \hat{P}$. Multiplying by $\rho_0$, taking the trace, and pulling the derivative outside yields:
\begin{gather}
    \text{Tr}[ \hat{P} \hat{D}(s) \rho_0 ] = \frac{i}{2\sigma} \frac{\partial}{\partial s} \text{Tr}[ \hat{D}(s) \rho_0 ] = \frac{i}{2\sigma} \frac{\partial}{\partial s} \left( e^{-\Gamma s^2} \right)  \nonumber \\
    = -i \frac{\Gamma s}{\sigma} e^{-\Gamma s^2}.
\end{gather}
Substituting this solution back into Eq.~\eqref{eq:P_offdiag_split}, we obtain the off-diagonal contribution:
\begin{equation}
    \text{Tr}(\hat{P} \rho_{+-}) = e^{-isb} \left( P_0 - i \frac{\Gamma s}{\sigma} \right) e^{-\Gamma s^2}.
\end{equation}
By symmetry, the conjugate cross-term yields $\text{Tr}(\hat{P} \rho_{-+}) = e^{isb} \left( P_0 + i \frac{\Gamma s}{\sigma} \right) e^{-\Gamma s^2}$.

\subsection{Total Mean Momentum Shift}

The final mean momentum $\langle P \rangle_f$ is the sum of these traces, weighted by the normalized coefficients $\mathcal{W}_{ij}$ of the conditioned state. Factoring out $P_0$ allows us to group the terms:
\begin{gather}
    \langle P \rangle_f = P_0 \Big[ \mathcal{W}_{++} + \mathcal{W}_{--} + \mathcal{W}_{+-} e^{-2isb} e^{-\Gamma s^2} + \mathcal{W}_{-+} e^{i2sb} e^{-\Gamma s^2} \Big] \nonumber \\
    - i \frac{\Gamma s}{\sigma} e^{-\Gamma s^2} \Big[ \mathcal{W}_{+-} e^{-2isb} - \mathcal{W}_{-+} e^{2isb} \Big].
\end{gather}
The expression inside the first bracket is precisely the trace of the normalized conditioned state $\rho_M^{PS}$, which is identically $1$. This perfectly isolates $P_0$. To find the net momentum shift $\Delta P$, we subtract this initial momentum $P_0$.

Using that $\mathcal{W}_{-+} = \mathcal{W}_{+-}^*$, the remaining term inside the second bracket takes the form $z - z^* = 2i\text{Im}(z)$. This cancels the imaginary $-i$ coefficient, yielding the final, strictly real expression for the anomalous momentum shift:
\begin{equation}
    \Delta P = 2 \frac{\Gamma s}{\sigma} e^{-\Gamma s^2} \text{Im} \Big( \mathcal{W}_{+-} e^{-2isb} \Big).
\end{equation}
This shows that while the position shift is driven by the diagonal elements, the momentum shift is a purely quantum phenomenon driven exclusively by the off-diagonal coherence terms, explicitly modulated by the initial momentum $b$ in the phase $e^{-ibs}$. In analogy with the position shift, we can factor this result against the characteristic momentum scale of the interaction, $s/\sigma$, to express the mean momentum shift as $\Delta P = (s/\sigma) \mathcal{A}_P(s)$. Thus, we define the dimensionless momentum amplification factor:
\begin{equation}
    \mathcal{A}_P(s) \equiv 2 \Gamma e^{-\Gamma s^2} \text{Im} \Big( \mathcal{W}_{+-} e^{-2isb} \Big).
\end{equation}
Substituting the explicit GAD coefficients $c_j^{\pm}$ (see Appendix~\ref{app:GAD-coef}), we obtain:
\begin{equation} \label{eq:amplification_p_appx}
    \mathcal{A}_P(s) = \frac{1}{2 P_{\text{succ}}} \Big[ - \Gamma e^{-\Gamma s^2} \sqrt{1-\gamma} \sin\theta_i \sin\theta_f \sin(2bs + \phi_i - \phi_f) \Big].
\end{equation}

\FloatBarrier 

\bibliographystyle{apsrev4-2}
\bibliography{reference}

\end{document}